\documentclass[plb,twocolumn,superscriptaddress]{revtex4}
\usepackage{epsfig}
\usepackage{graphicx}
\usepackage{palatino}
\usepackage[english]{babel}
\usepackage{hyphenat}
\usepackage{amsmath}
\usepackage{amssymb}
\usepackage{slashed}
\usepackage{epstopdf}
\usepackage{xcolor}
\usepackage{booktabs}
\definecolor{lcolor}{rgb}{0.,0.0,0.}
\definecolor{citcolor}{rgb}{0,0.,0.5}
\usepackage[breaklinks,colorlinks,urlcolor=blue,citecolor=blue,linkcolor=blue]{hyperref}
\usepackage{hhline}
\usepackage{multirow}
\usepackage{ltablex}

\newcommand{\beq}{\begin{equation}}
\newcommand{\eeq}{\end{equation}}
\newcommand{\bea}{\begin{eqnarray}}
\newcommand{\eea}{\end{eqnarray}}
\newcommand{\dis}{\displaystyle}

\newcommand{\bem}{\begin{multline}}
\newcommand{\eem}{\end{multline}}
\newcommand{\beg}{\begin{gather}}
\newcommand{\eeg}{\end{gather}}

\newcommand{\nn}{\nonumber}

\newcommand{\ben}{\begin{eqnarray*}}
\newcommand{\een}{\end{eqnarray*}}

\newcommand{\eq}[1]{\begin{align}#1\end{align}}
\begin{document}
\title{{\bf Symmetric cumulants as a probe of the proton substructure at LHC energies \\}}
\author{Javier L. Albacete}
\email[]{albacete@ugr.es}
\affiliation{CAFPE and Departamento de F\'isica Te\'orica y del Cosmos,  Universidad de Granada, E-18071 Campus de Fuentenueva, Granada, Spain.}
\author{Hannah Petersen}
\email[]{petersen@fias.uni-frankfurt.de}
\affiliation{Frankfurt Institute for Advanced Studies, Ruth-Moufang-Strasse 1, 60438 Frankfurt am Main, Germany.}
\affiliation{Institute for Theoretical Physics, Goethe University,Max-von-Laue-Strasse 1, 60438 Frankfurt am Main, Germany.}
\affiliation{GSI Helmholtzzentrum f{\"u}r Schwerionenforschung, Planckstr. 1, 64291 Darmstadt, Germany.}
\author{Alba Soto-Ontoso}
\email[]{ontoso@fias.uni-frankfurt.de}
\affiliation{CAFPE and Departamento de F\'isica Te\'orica y del Cosmos,  Universidad de Granada, E-18071 Campus de Fuentenueva, Granada, Spain.}
\affiliation{Frankfurt Institute for Advanced Studies, Ruth-Moufang-Strasse 1, 60438 Frankfurt am Main, Germany.}

\begin{abstract}
We present a systematic study of the normalized symmetric cumulants, NSC(n,m), at the eccentricity level in proton-proton interactions at $\sqrt s\!=\! 13$~TeV within a wounded hot spot approach.
We focus our attention on the influence of spatial correlations between the proton constituents, in our case gluonic hot spots, on this observable. We notice that the presence of short-range repulsive correlations between the hot spots systematically decreases the values of NSC(2,3) and NSC(2,4) in mid- to ultra-central collisions while increases them in peripheral interactions. In the case of NSC(2,3) we find that, as suggested by data, an anti-correlation of $\varepsilon_2$ and $\varepsilon_3$ in ultra-central collisions, i.e. NSC(2,3) $<0$, is possible within the correlated scenario while it never occurs without correlations when the number of gluonic hot spots is set to three. We attribute this fact to the decisive role of correlations on enlarging the probability of interaction topologies that reduce the value of NSC(2,3) and, eventually, make it negative. Further, we explore the dependence of our conclusions on the number of hot spots, the values of the hot spot radius and the repulsive core distance. Our results add evidence to the idea that considering spatial correlations between the subnucleonic degrees of freedom of the proton may have a strong impact on the initial state properties of proton-proton interactions \cite{Albacete:2016gxu}.    
\end{abstract}

\maketitle
\section{Introduction}
The vast amount of high-precision data collected in the recent years at the Large Hadron Collider (LHC) has allowed to go beyond the analysis of event-averaged observables and explore higher order moments of their probability distributions. In the context of heavy ion collisions, a very recent tool to study the properties of the Quark Gluon Plasma (QGP) are the correlations between different flow harmonics $v_{n}$ i.e. the symmetric cumulants defined as \cite{Bilandzic:2013kga,DiFrancesco:2016srj}
\eq{
{\rm SC(n,m)}=\langle v_n^2 v_m^2\rangle-\langle v_n^2\rangle \langle v_m^2\rangle
\label{scnm}
}
or in their normalized version 
\eq{
{\rm NSC(n,m)}=\displaystyle\frac{\langle v_n^2 v_m^2\rangle-\langle v_n^2\rangle\langle v_m^2\rangle}{\langle v_n^2\rangle \langle v_m^2\rangle}
\label{nscnm}
}
that eliminate the dependence on the absolute magnitude of $v_{n(m)}$.
The measurement of Eq.~\ref{scnm} would be zero by definition, if the fluctuations of $v_n$ and $v_m$ were totally uncorrelated, in the same way as the Pearson's correlation coefficient. Instead, a positive value of Eq.~\ref{scnm} implies that an event with $v_n>\langle v_n\rangle$ would be more likely to have $v_m>\langle v_m\rangle$.
In particular, by measuring SC(2,3) we can gain information about initial state fluctuations whereas SC(2,4) is mostly sensitive to the strongly interacting medium properties \cite{Giacalone:2016afq}.  

The experimental study of symmetric cumulants was pioneered by the ALICE Collaboration \cite{ALICE:2016kpq} in Pb+Pb collisions at $\sqrt {s_{\rm NN}}\!=\! 2.76$~TeV. Recently, the CMS Collaboration has performed an experimental analysis of the symmetric cumulants as a function of the multiplicity in the three collision systems available at the LHC: p+p, p+Pb and Pb+Pb \cite{CMS:2017saf}. The experimental results suggest a similar pattern across systems. Concretely, SC(2,4) is always positive although its multiplicity dependence varies from p+p to Pb+Pb. Further, NSC(2,4) is clearly modified when varying the system size. On the contrary, the sign of SC(2,3) is strongly multiplicity-dependent: at low multiplicities SC(2,3) is found to be positive. However, it turns out to be negative for very high multiplicities, $N_{\rm trk}^{\rm offline}>60$ in both p+Pb and Pb+Pb and $N_{\rm trk}^{\rm offline}\sim 100$ in p+p. Moreover, NSC(2,3) in the high-multiplicity regime is found to have not only the same sign in the three collision systems but the same quantitative value as well. It should be noted that in the p+p case although the trend of NSC(2,3) to smaller values when increasing the multiplicity is visible, the systematic uncertainties make it compatible with zero. Relating the negative sign of NSC(2,3) at very high multiplicities with the initial geometry of the proton is the main goal of this work. 

Previous to the symmetric cumulants study the similarities between high-multiplicity proton-proton interactions with p+Pb and Pb+Pb were already observed in the individual flow harmonic coefficients \cite{Khachatryan:2016txc,Aad:2015gqa,Khachatryan:2010gv} and the enhanced production of multi-strange hadrons measurements \cite{ALICE:2017jyt}. All together, the experimental data is constantly reigniting the debate on whether collective effects, precedently attributed to the formation of QGP droplets, are being observed in small collision systems such as p+p and p+Pb \cite{Schlichting:2016sqo,Weller:2017tsr,dEnterria:2010xip,Dusling:2015gta}. 

From a theoretical point of view, the well established paradigm of either Monte Carlo Glauber \cite{Giacalone:2016afq} or Color Glass Condensate \cite{Eskola:2017imo,Gardim:2016nrr,Niemi:2015qia} initial conditions followed by viscous hydrodynamic evolution has successfully described the data on SC(n,m) in the Pb+Pb case. Turning to smaller systems, the situation is less conclusive. On the one hand, there have been attempts to describe the values of SC(n,m) in p+Pb by computing them in terms of eccentricities, i.e. replacing $v_{n(m)}$ by $\varepsilon_{n(m)}$ in Eqs.~\ref{scnm}-\ref{nscnm}, within wounded quark models. These studies lead to the correct negative sign of SC(2,3) at high multiplicities but the magnitude is off \cite{Broniowski:2016pvx,talk2,Welsh:2016siu}. Also without hydrodynamic evolution, qualitative agreement with the p+Pb results has been achieved in an initial state model where the partons in the projectile coherently scatter off the color fields in the heavy nuclear target \cite{Dusling:2017aot,Dusling:2017dqg}. Up to today we are unaware of any theoretical prediction for the values of SC(n,m) in p+p interactions at LHC energies, although results for RHIC energies were presented in \cite{Broniowski:2016pvx}.

A key ingredient in the computation of the symmetric cumulants in any of these frameworks is the parametrization of the initial geometry of the collision. Especially the smaller systems exhibit a high degree of sensitivity to the description of the proton structure and its fluctuations. The importance of considering subnucleonic degrees of freedom when describing the elliptic flow in p+Pb within the IP-Glasma framework was realized in \cite{Mantysaari:2017cni}. In parallel, considering the proton constituents to be subjected to spatial correlations has been shown to have a substantial impact on the values of $\varepsilon_2$ and $\varepsilon_3$ \cite{Albacete:2016gxu} beside other features of hadronic interactions such as the hollowness effect \cite{RuizArriola:2016ihz,Albacete:2016pmp}. This work constitutes the natural extension of our previous studies on the initial state properties of proton-proton collisions in terms of eccentricities \cite{Albacete:2016gxu} by exploring not only their mean but their fluctuations. As we shall explain along the manuscript we rely, for simplicity, on a geometrical picture of the collision. Therefore we use Monte Carlo Glauber \cite{Glauber} simulations, where the proton is composed by, in principle, 3 gluonic hot spots. By relating the hot spots to the gluon clouds radiated by the valence quarks considering the proton to be formed by 3 constituents becomes natural. However, the possibility of having a different number, $N_{hs}$, that may account for other elements such as the large-$x$ sea quarks is explored in this work. The centrality selection is done in terms of the entropy deposition as a proxy of particle production. We find that the inclusion of short-range repulsive correlations has a critical impact on the sign of NSC(2,3) in ultra central collisions. The net effect of the presence of correlations is to reduce the value of NSC(2,3) with respect to the uncorrelated scenario in the more central collisions and even push it to negative values in the highest centrality bins. An intuitive interpretation of this result is given by characterizing the topology of the interaction in terms of the number of wounded hot spots and of collisions between them. In the case of NSC(2,4), the results are qualitatively the same as for NSC(2,3) although it always remains positive within the regions of the parameter space explored in this work. In order to disentangle the possible phenomena that could contribute to the negative sign of NSC(2,3) we compute it for different values of the parameter space. We conclude that, as expected, not only the presence of correlations is important but also the interplay between the different scales of the problem, that is, the number of gluonic hot spots, their radius and the value of the repulsive distance. The radius of the hot spot, $R_{hs}$, has been studied, in terms of the correlation length of the gluon field strengths inside hadrons, via lattice QCD calculations \cite{DiGiacomo:1992hhp} and within perturbative \cite{PhysRevD66034031} and non-perturbative \cite{Schafer:1996wv} frameworks. On the contrary the value of the repulsive distance, $r_c$, apart from being different from zero \cite{Albacete:2016pmp}, is essentially unconstrained. So it is the number of gluonic hot spots. Thus, this study guided by the experimental data on NSC(n,m) helps to restrict the values of $N_{hs}$, $R_{hs}$ and $r_c$ within our model. 

The organization of the paper is as follow. We begin by reviewing the main ingredients of our model in Sec.~\ref{model}. Then, in Sec.~\ref{results_important}, the results for NSC(2,3) and NSC(2,4) as a function of centrality are presented. As the most interesting results occur on the 0-1\% centrality bin, we focus on it in Sec.~\ref{ultracentral} and study the role of the interaction topology together with a scan of the parameter space. Further, we study the sensitivity of our results to the number of constituents in each proton in Sec.~\ref{nhs}. Finally, our conclusions and future lines of work are given in Sec.~\ref{final}.
 
\section{Wounded hot spots model}
\label{model}
In this section, we briefly expose the main ingredients of our Monte Carlo Glauber event generator for proton-proton interactions that follows similar steps than others in the literature \cite{Broniowski:2007nz,Loizides:2014vua}. For a detailed description of the model see \cite{Albacete:2016gxu}. In the following, a proton is considered to be formed by 3 hot spots. The comparison of our results for $N_{hs}\!=\!(2,4)$ are given in Sec.~\ref{nhs}.

In each p+p event, after generating a random impact parameter for the collision, we sample the transverse positions of the three hot spots in each proton $\lbrace \vec s_i \rbrace$ according to the distribution 

\eq{
D(\vec s_{1},\vec s_{2},\vec s_{3})&=C \dis\prod_{i=1}^3e^{- s_i^2/R^2}\delta^{(2)}(\vec{s}_1+\vec{s}_2+\vec{s}_3)
\times \nn \\
&\dis\prod_{\substack{{i<j}\\{i,j=1}}}^3\left(1-e^{-\mu\vert\vec{s}_i-\vec{s}_j\vert^2/R^2}\right).
\label{corr}
}

where $C$ is a normalization constant and $R$ is the average radius. It should be noted that the extension of Eq.~\ref{corr} to an arbitrary number of hot spots is direct. Most of the models in the literature \cite{Bernhard:2016tnd,Bozek:2016kpf,Welsh:2016siu,Mantysaari:2017cni,Weller:2017tsr,Loizides:2016djv,Mitchell:2016jio} implement a proton geometry following the two first terms of Eq.~\ref{corr} i.e. the hot spots are distributed according to Gaussian functions with the natural constraint of fixing the centre of mass of the constituents system to the centre of the proton. However, with this set up the most probable configurations are the ones with three hot spots in the middle of the proton and the one with two hot spots fully or partially overlapping and the third one separated due to the $\delta$-function like in the quark-diquark model. The third term of Eq.~\ref{corr} allows us to go beyond these approaches by implementing short range repulsive correlations among all pairs of hot spots that effectively enlarge the mean transverse separation $\vert \vec s_{i}-\vec s_{j}\vert$ between them. The size of this correlation is controlled by an effective repulsive core distance $r_c^2\equiv R^2/\mu$. The original motivation to consider these additional spatial correlations within our model was their critical impact on the dynamical explanation of the hollowness effect \cite{Albacete:2016pmp}. Although it constitutes the main novelty of our phenomenological model with respect to others in the literature in this context it should be noted that the necessity of spatial correlations has been already entertained in the nuclear case \cite{Denicol:2014ywa,Alvioli:2009ab,Blaizot:2014wba}. Further, a similar mechanism prevents the ropes in the DIPSY event generator to be in a highly energetic color state \cite{Bierlich:2014xba}. All along the manuscript we will focus on comparing the results obtained in the uncorrelated scenario ($\mu\to\infty$) with the correlated case.

Once the hot spots are located in both target and projectile, the next step in the Monte Carlo simulation is to decide which of them have been wounded \cite{Bialas:1977en,Bialas:1976ed} i.e. have collided at least once. Our collision criterion consists of sampling the inelasticity density 
\eq{
G_{\rm{in}}(d)&=2e^{-d^2/2R_{hs}^2}-(1+\rho_{hs}^2)e^{-d^2/R_{hs}^2}
\label{gin}
}
that depends on the radius of the hot spot $R_{hs}$, the transverse distance between the pair of hot spots considered $d$ and the ratio of real and imaginary parts of the hot spot-hot spot scattering amplitude $\rho_{hs}$. Thus, in each event, the maximum number of wounded hot spots $N_{w}$ and collisions $N_{\rm coll}$ is 6 and 9 respectively. In Section~\ref{topology} we will use these two variables to characterize interaction topologies. 

Subsequently, we consider that each wounded hot spot located at ($x_w, y_w$) deposits a random amount of entropy according to

\eq{s(x,y)=s_0\dis\frac{1}{\pi R_{hs}^2}\exp\left(-\dis\frac{(x-x_w)^2+(y-y_w)^2}{R_{hs}^2}\right)
\label{entropy}
}
 
where $s_0$ fluctuates independently for each wounded hot spot following a double Gamma distribution where parameters, given in Table~\ref{param_entropy}, are fixed by the assumption that entropy deposition is related to particle production \cite{Albacete:2016gxu}. Moreover, the centrality classes considered ($[0\!-\!0.1\%],[0.1\!-\!1\%], [1\!-\!5\%],[5\!-\!10\%],[10\!-\!20\%],[20\!-\!30\%]\ldots[90\!-\!100\%]$) are defined via the entropy deposition.

At this point, all the wounded hot spots contribute to the calculation of the spatial eccentricity moments that characterize the initial geometry anisotropy of the collision

\eq{\varepsilon_n=\dis\frac{\sqrt{\langle\dis\sum_{i=1}^{N_w} r_{i}^{n}\cos(n\phi_i)\rangle^2+\langle\dis\sum_{i=1}^{N_w} r_{i}^{n}\sin(n\phi_i)\rangle^2}}{\langle\dis\sum_{i=1}^{N_w} r_{i}^{n}\rangle}
\label{ecc}
}
where $\langle \cdot \rangle$ in Eq.~\ref{ecc} denotes the average weighted by the entropy deposition given by Eq.~\ref{entropy}. We compute Eq.~\ref{ecc} in the participant plane on an event-by-event basis.

Regarding the model parameters, by default we use the same values as the ones from \cite{Albacete:2016gxu} at $\sqrt s\!=\!13$~TeV given in Table~\ref{param}. However, in Section \ref{param_space} we extend our calculation to other regions of the parameter space. The three correlation scenarios under consideration in this work are the following: first, $r_c\!=\! 0.4$ refers to the correlated scenario with $\lbrace R_{hs}$, $R$, $\rho_{hs}\rbrace$ constrained to reproduce the extrapolated values of the total cross section and the ratio of real and imaginary parts of the scattering amplitude \cite{Cudell:2002xe}. In the second case we impose the latter constraints to $\lbrace R_{hs}$, $R$, $\rho_{hs}\rbrace$ after setting $r_c\!=\!0$. However, in order to do a more realistic comparison between the correlated and uncorrelated scenarios it is necessary to avoid the intrinsic swelling effects due to the presence of repulsive correlations. The case labeled as "$\langle s_1 \rangle$ fixed" constitutes an attempt to perform this task by fixing the r.m.s of the spatial probability distribution given by Eq.~\ref{corr} to be the same as in the $r_c\!=\! 0.4$ scenario. In the following plots the uncorrelated results will be exhibited as a band bounded by the $r_c\!=\!0$ and $\langle s_1 \rangle$ fixed cases to display the different possibilities considered. 

Once the building blocks of our model have been presented in the next sections we display its results for the normalized symmetric cumulants given by Eq.~\ref{nscnm} obtained after generating 4.5 million events. Only the events with at least two wounded hot spots contribute to the averages in the following plots.    

\begin{table}
\begin{center}
\begin{tabular}{|c|*{7}{c|}}
 \multicolumn{3}{|c|} {$\bf{r_c\!=\!0.4}$~{\bf fm}} & \multicolumn{3}{|c|}{$\bf{r_c\!=\!0}$} 
  & \multicolumn{1}{|c|} {$\bf{\langle s_1 \rangle}$ \bf fixed} \\ \toprule[0.5mm]
 $R_{hs}$~[fm]&$R$~[fm] & $R_p$~[fm] & $R_{hs}$~[fm]&$R$~[fm] &$R_p$~[fm] & $R$~[fm]  \\ \toprule[0.5mm]
 0.32&0.76 & 1.34 & 0.41&0.75 & 1.28 &0.87
\end{tabular}
\caption{Default values of the parameters characterizing the hot spots distribution Eq.~\ref{corr} and their probability to interact Eq.~\ref{gin}. We set $\rho_{hs}\!=\!0.1$ in all cases. On the last column, the values of $R$ for the "$\langle s_1 \rangle$ fixed" case are shown.}
\label{param}
\end{center}
\end{table}

\begin{table}
\begin{center}
\begin{tabular}{c|c|*{6}{c|}} 
&  $\overline n_1$&$\kappa_1$ & $\overline n_2$  & $\kappa_2$ & $\alpha$  \\ \toprule[0.5mm]
 $\bf{r_c\!=\!0.4}$~{\bf fm} & 27.29 & 1.51 & 4.68 & 1.66 & 0.37\\ \toprule[0.5mm]
 $\bf{r_c\!=\!0}$& 26.26 & 1.55 & 4.16 & 1.79 & 0.31\\ \toprule[0.5mm]
 $\bf{\langle s_1 \rangle}$ \bf fixed & 26.47 & 1.42 & 4.54 & 1.73 & 0.37
\end{tabular}
\caption{Default values of the parameters of the double Gamma distribution that characterizes the fluctuating amount of entropy each wounded hot spot deposits, $s_0$.}
\label{param_entropy}
\end{center}
\end{table}

\section{Normalized symmetric cumulants vs centrality}
\label{results_important}
\begin{figure}[htb]
\begin{center}
\includegraphics[scale=0.465]{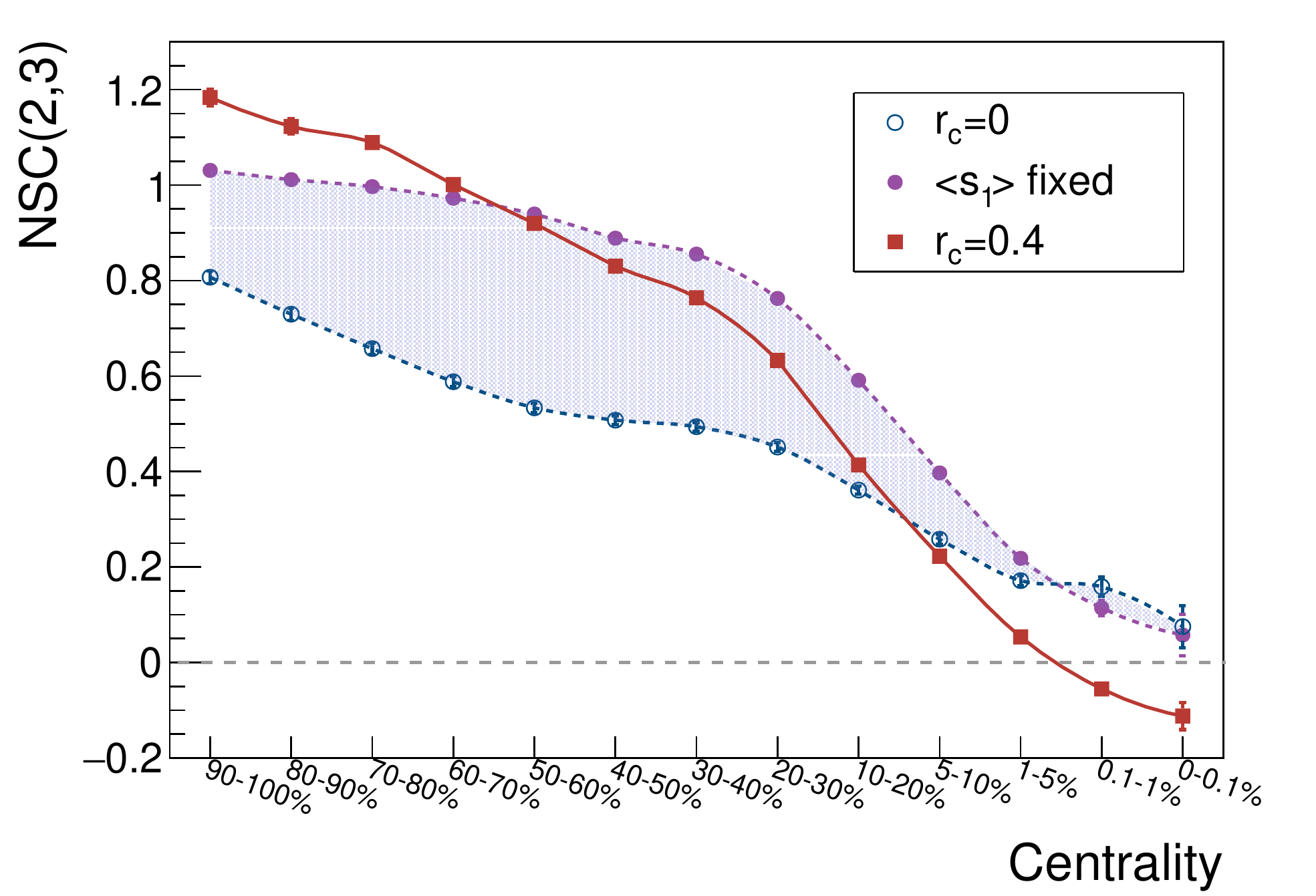} 
\end{center}
\vspace*{-0.5cm}
\caption[a]{Average value of NSC(2,3) as a function of the centrality range for $r_c\!=\!0$ (blue short-dashed line connecting open blue circles), $\langle s_1 \rangle$ fixed (purple short-dashed line connecting filled purple circles) and $r_c\!=\!0.4$~fm (red solid line connecting filled red squares). The error bars represent statistical uncertainties while the light violet band indicates the theoretical uncertainty associated to the choice of parameters that define the uncorrelated scenario.}
\label{nsc23_centrality}
\end{figure}

The most important result of this paper is shown in Fig.~\ref{nsc23_centrality} where we represent the event-averaged value of NSC(2,3) as a function of centrality. A common feature in the three correlation scenarios is the fact that NSC(2,3) decreases from peripheral to central collisions as suggested by data. Focusing on the effect of the short-range repulsive correlations we observe how they enlarge the positive correlation of $\varepsilon_2$ and $\varepsilon_3$ in the peripheral regime. However, their repercussion in the very central collisions is precisely the opposite. Finally, the most striking effect of the spatial correlations is observed in the ultra-central bins [0-0.1\%] and [0.1-1\%]: only in the $r_c\!=\!0.4$ case there exists an anti-correlation of $\varepsilon_2$ and $\varepsilon_3$ as data dictates. Then, we conclude that the experimental evidence of NSC(2,3)$<0$ may back up the necessity to consider correlated proton constituents. 
 
\begin{figure}[htb]
\begin{center}
\includegraphics[scale=0.465]{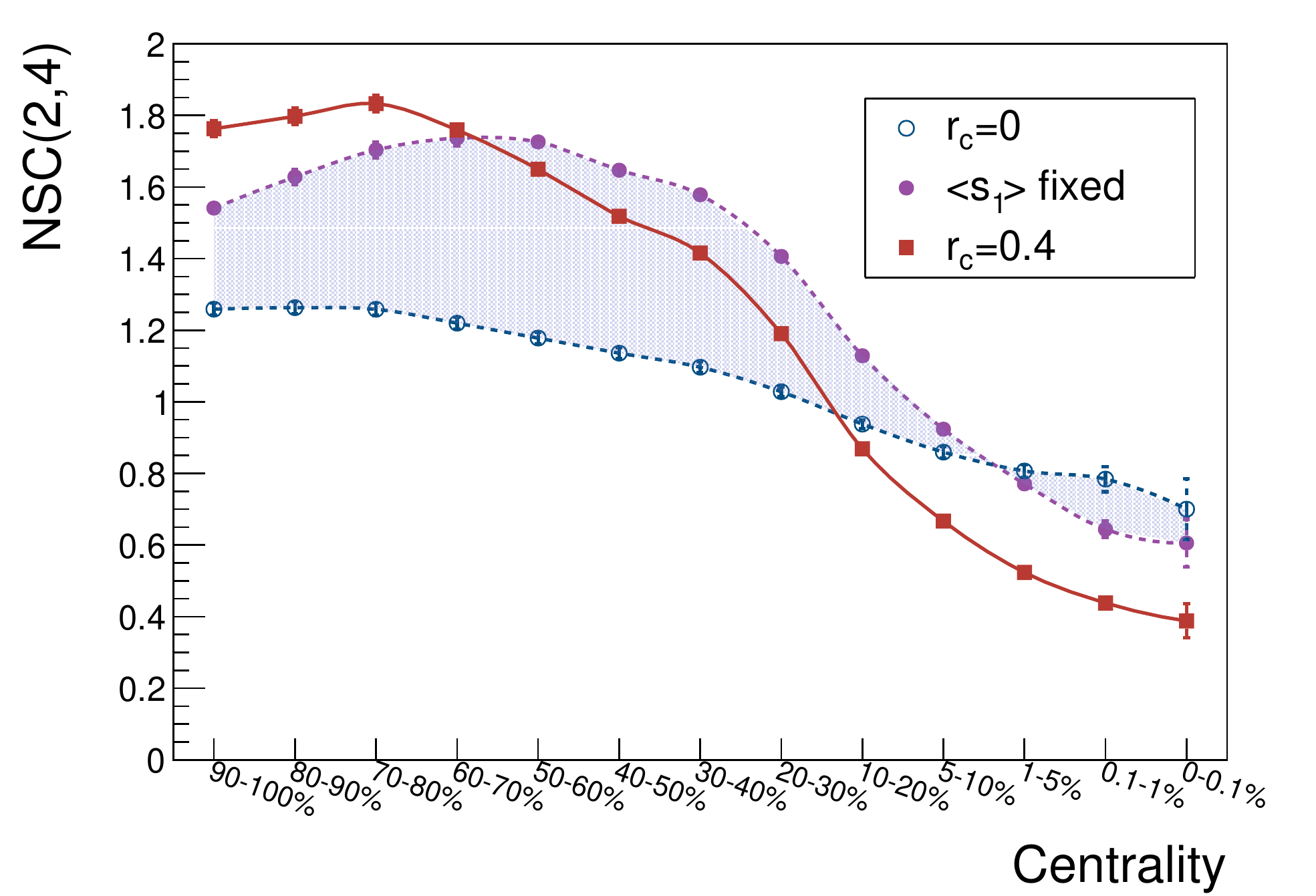} 
\end{center}
\vspace*{-0.5cm}
\caption[a]{Average value of NSC(2,4) as a function of the centrality range for $r_c\!=\!0$ (blue short-dashed line connecting blue open circles), $\langle s_1 \rangle$ fixed (purple short-dashed line connecting filled purple circles) and $r_c\!=\!0.4$~fm (red solid line connecting filled red squares). The error bars represent statistical uncertainties while the light violet band indicates the theoretical uncertainty associated to the choice of parameters that define the uncorrelated scenario.}
\label{nsc24_centrality}
\end{figure}

An important comment is in order at this point. A direct comparison with the experimental data is not straightforward especially in the low multiplicity regime where non-flow dijet contributions, totally absent in our initial state coordinate space approach, dominate the measured values of NSC(n,m). Another important issue is to ensure that the centrality bin selection is exactly the same both in our approach based on the entropy deposition and in the data in terms of $N_{\rm {trk}}^{\rm {offline}}$ \cite{Khachatryan:2016txc}. Extending our calculation to higher centrality bins is doable but computationally expensive. However, a precise theory-to-data comparison, although desirable, is not the main objective of this work. Our purpose is to present, for the first time in the literature, a particular mechanism i.e. the presence of spatial correlations inside the proton that builds up a negative value of NSC(2,3) in the highest centrality bin at the geometric level. 

In the case of NSC(2,4), the role of the repulsive correlations is qualitatively the same as in the NSC(2,3) calculation: in peripheral collisions the value of NSC(2,4) is larger in the $r_c\!=\!0.4$ case than in the uncorrelated scenarios and the situation gets reversed at barely the same centrality bin. As well, we find the absolute value of NSC(2,4) to be larger than NSC(2,3) in all the centrality bins as it is the case in the data. We would also like to remark that in our approach the symmetric cumulants are almost flat in the mid-to-peripheral interactions but thanks to a dissection of the very central bins we see a clear centrality dependence. This is consistent with our previous calculations of the average values of the spatial eccentricity moments \cite{Albacete:2016gxu}.

A geometric and intuitive interpretation of the fact that only in the correlated case NSC(2,3)$<0$ in the [0-1\%] centrality bin is given in the following section. It should be noted that, for this purpose, we have merged the two highest centrality bins, [0-0.1\%] and [0.1-1\%], into a single one in order to improve the statistics. 
\section{Ultra-central events}
\label{ultracentral}
All the results presented in this Section refer to the [0-1\%] centrality bin. We restrict our calculations to this bin because as we have emphasized in the previous section, we are interested in the change of sign of NSC(2,3).  
\subsection{Role of the interaction topology}
\label{topology}
In order to capture the effect of the spatial correlations we characterize each proton-proton interaction by its number of wounded hot spots and the number of collisions ($N_w,N_{\rm coll}$), the two basic quantities of any Monte-Carlo Glauber calculation. We dub each ($N_w,N_{\rm coll}$)-configuration as \textit{interaction topology}. In our case, given that we consider the proton to be formed by three hot spots $N_w\in[2,6]$ and $N_{\rm coll}\in[1,9]$. 

\begin{figure}[htb]
\begin{center}
\includegraphics[scale=0.465]{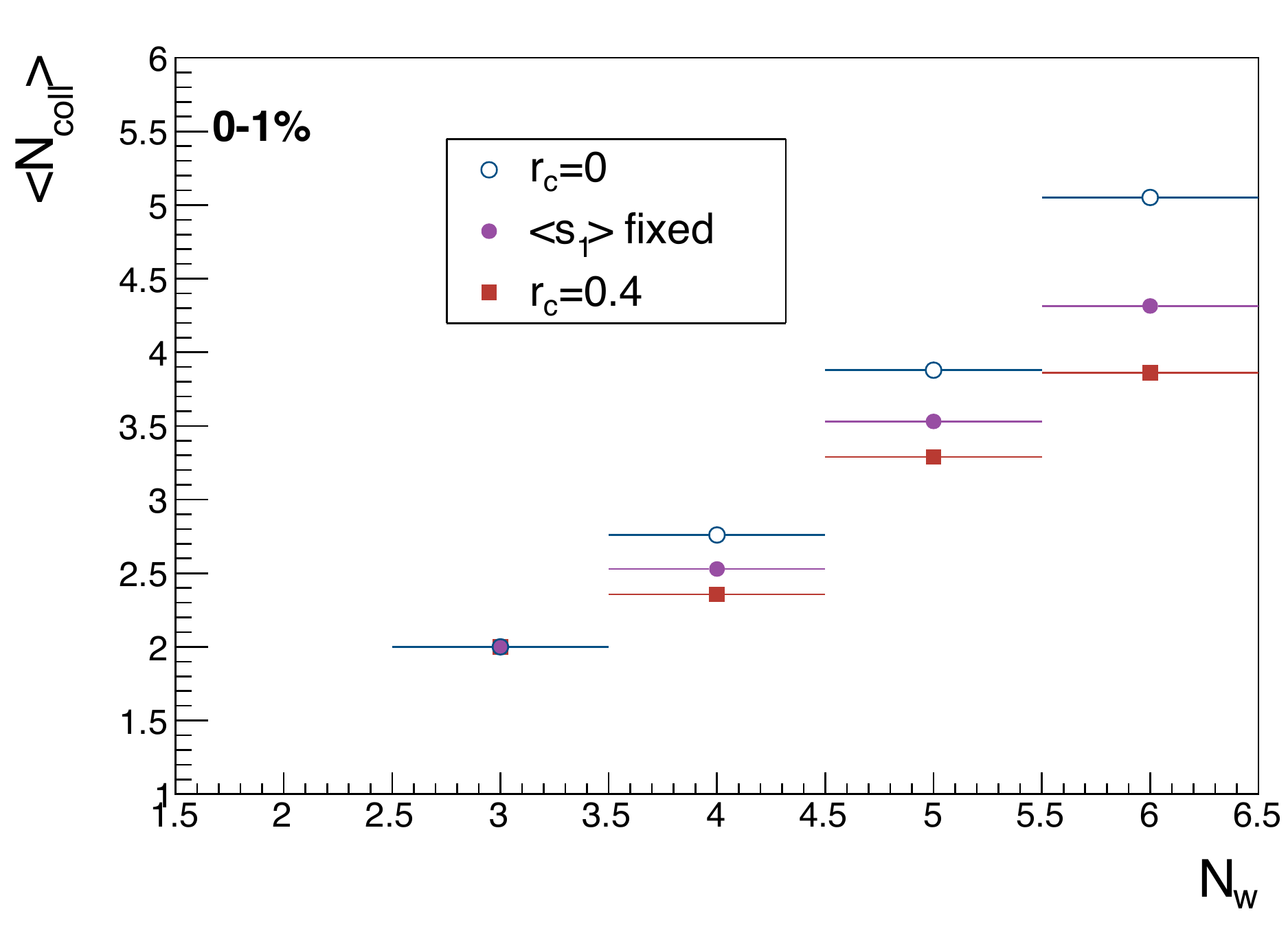} 
\end{center}
\vspace*{-0.5cm}
\caption[a]{Average number of collisions as a function of the number of wounded hot spots for $r_c\!=\!0$ (open blue circles), $\langle s_1 \rangle$ fixed (filled purple circles) and $r_c\!=\!0.4$~fm (filled red squares). }
\label{ncoll_nw}
\end{figure}

\begin{figure}[htb]
\begin{center}
\includegraphics[scale=0.6]{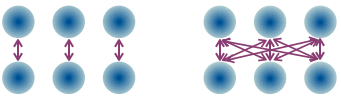} 
\end{center}
\vspace*{-0.5cm}
\caption[a]{Sketch representing the interaction topologies preferred in the correlated case (left) and in the uncorrelated one (right). The purple arrows represent the collisions between the hot spots.}
\label{int_topology}
\end{figure}

We begin our analysis by computing the average number of collisions as a function of the number of wounded hot spots for the three different scenarios introduced above. The results are shown in Fig.~\ref{ncoll_nw}. First of all, as we describe the entropy deposition in an incoherent way i.e. on average the more wounded hot spots the more entropy is deposited, the configurations in which only two hot spots collide cannot create enough entropy to be part of the [0-1\%] centrality bin. Then, the minimum number of wounded hot spots is three and, in this case, $\langle N_{\rm coll}\rangle\!=\!2$ in all the correlation scenarios as it is the only existing configuration. However, for $N_w>3$ the average number of collisions starts to differ between the three different cases. We observe that $\langle N_{\rm coll}\rangle$ is systematically reduced when including repulsive correlations with respect to the uncorrelated cases. This effect has a very straightforward interpretation: enlarging the mean transverse distance between the hot spots reduces the probability of having interaction topologies with a high number of collisions. In other words, the repulsive correlations spread the hot spots in the transverse plane and as a consequence enhance the probability of the hot spots to collide by pairs over the configurations in which all hot spots in the projectile interact with all the others in the target, as it is schematically represented in Fig.~\ref{int_topology}.

To connect this fact with the total value of NSC(2,3) we would like to understand the individual contributions from the different interaction topologies. For this purpose we define a weighted version of NSC(n,m) denoted NSC$_{\rm w}$(n,m) as follows

\eq{{\rm NSC_w(n,m)}\equiv\mathcal{P}(N_{w})\cdot \mathcal{P}(N_w\vert N_{\rm coll}) \cdot {\rm NSC(n,m)}\Big\vert_{N_w,N_{\rm coll}}
\label{nscw}
}

where
\begin{itemize}
\item{} $\mathcal{P}(N_w)$ is the probability of having a certain number of wounded hot spots.
\item{} For a given $N_w$, $\mathcal{P}(N_w\vert N_{\rm coll})$ represents the probability of having a certain number of collisions between the hot spots. 
\item{} ${\rm NSC(n,m)}\Big\vert_{N_w,N_{\rm coll}}$ is the value of NSC(n,m) for each interaction topology. 
\end{itemize}
The error of NSC$_{\rm w}$(n,m)is computed by adding the statistical uncertainties of each term in Eq~\ref{nscw} in quadrature.
Essentially, by summing NSC$_{\rm w}$(n,m) over all the possible configurations ($N_w,N_{\rm coll}$) one recovers NSC(n,m). This new quantity allows us to decompose the value of NSC(2,3) and investigate the contribution of each interaction topology separately. From now on, to facilitate the discussion, we only show the comparison between $\langle s_1 \rangle$ fixed and $r_c\!=\!0.4$ scenarios. We have checked that the same conclusions as in the $\langle s_1 \rangle$ fixed case hold for $r_c\!=\!0$.

\begin{figure}[htb]
\begin{center}
\includegraphics[scale=0.465]{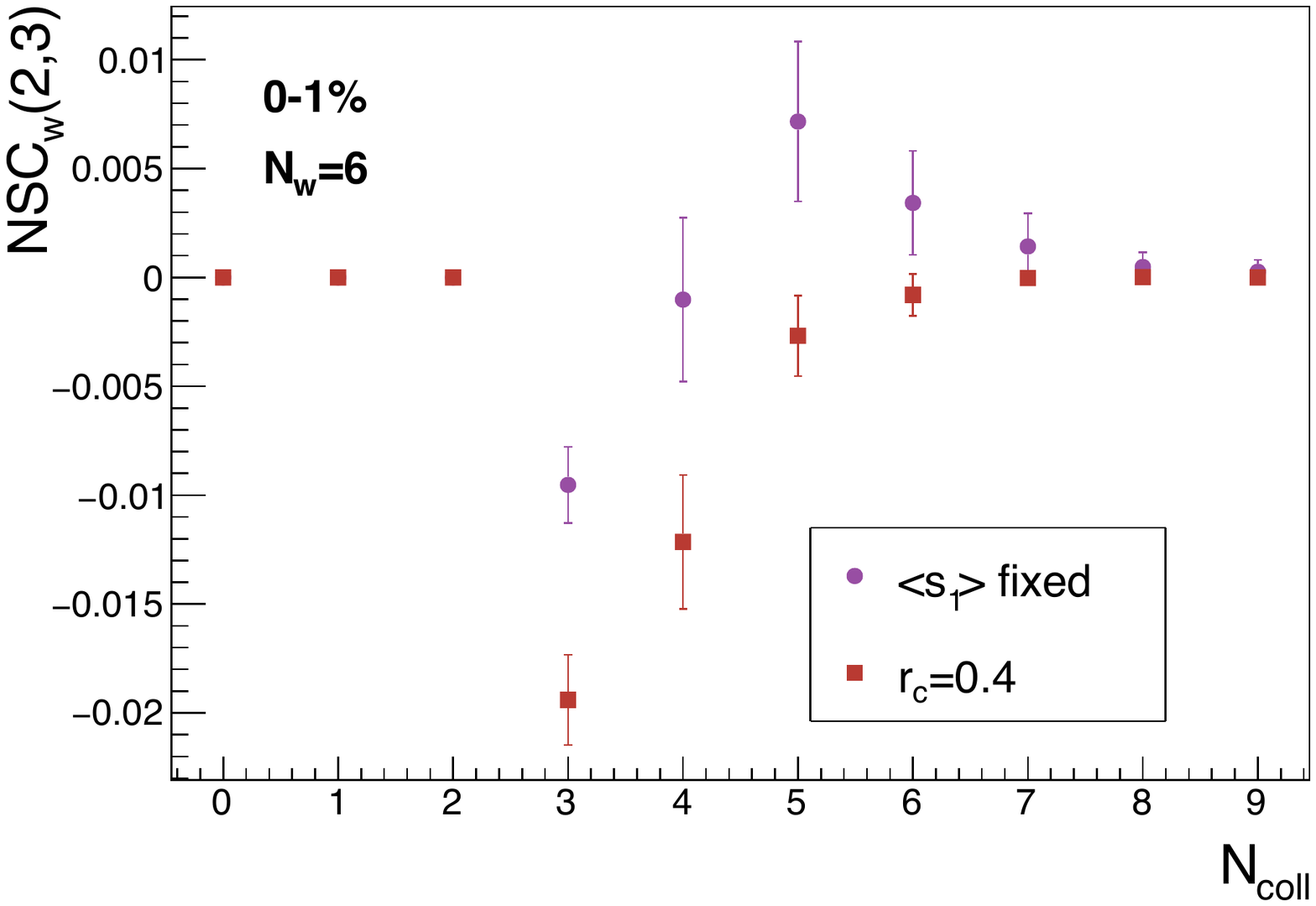} 
\end{center}
\vspace*{-0.5cm}
\caption[a]{Average value of NSC$_{\rm w}$(2,3) as a function of the number of collisions after selecting the events with $N_{\rm w}\!=\!6$ for $\langle s_1 \rangle$ fixed (filled purple circles) and $r_c\!=\!0.4$~fm (filled red squares).}
\label{nsc23_nw6}
\end{figure}

In Fig.~\ref{nsc23_nw6} we show a particular example of the output of our calculation for NSC$_{\rm w}$(2,3) by selecting the events with $N_w\!=\!6$. Two important results can be extracted from this figure. First, as already suggested by Fig.~\ref{ncoll_nw}, configurations with a large number of collisions, e.g. $N_{\rm coll}>6$, only occur in the uncorrelated case where the three hot spots are closer to each other or, equivalently, clustered. Second, and more important, the value of NSC$_{\rm w}$(2,3) shows a clear dependence on $N_{\rm coll}$: configurations with a smaller number of collisions reduce the value of NSC(2,3) and, eventually, contribute negatively. Then, in our picture, the inclusion of spatial correlations inside the proton modifies the weight of each interaction topology in such a way that these configurations are enhanced. This feature provides a natural explanation for the different sign of NSC(2,3) in the uncorrelated and correlated scenarios. 

\begin{figure}[htb]
\begin{center}
\includegraphics[scale=0.465]{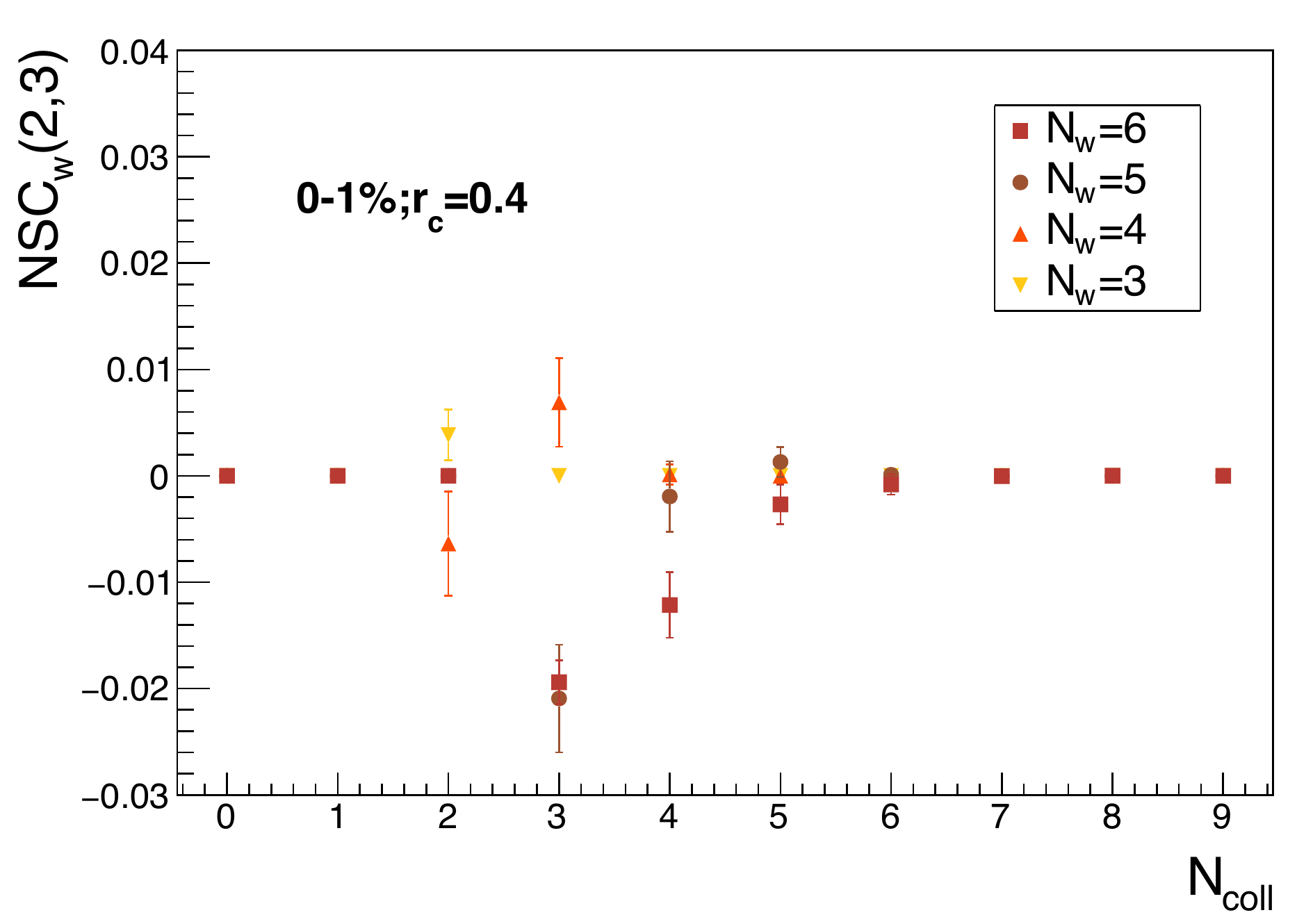} 
\includegraphics[scale=0.465]{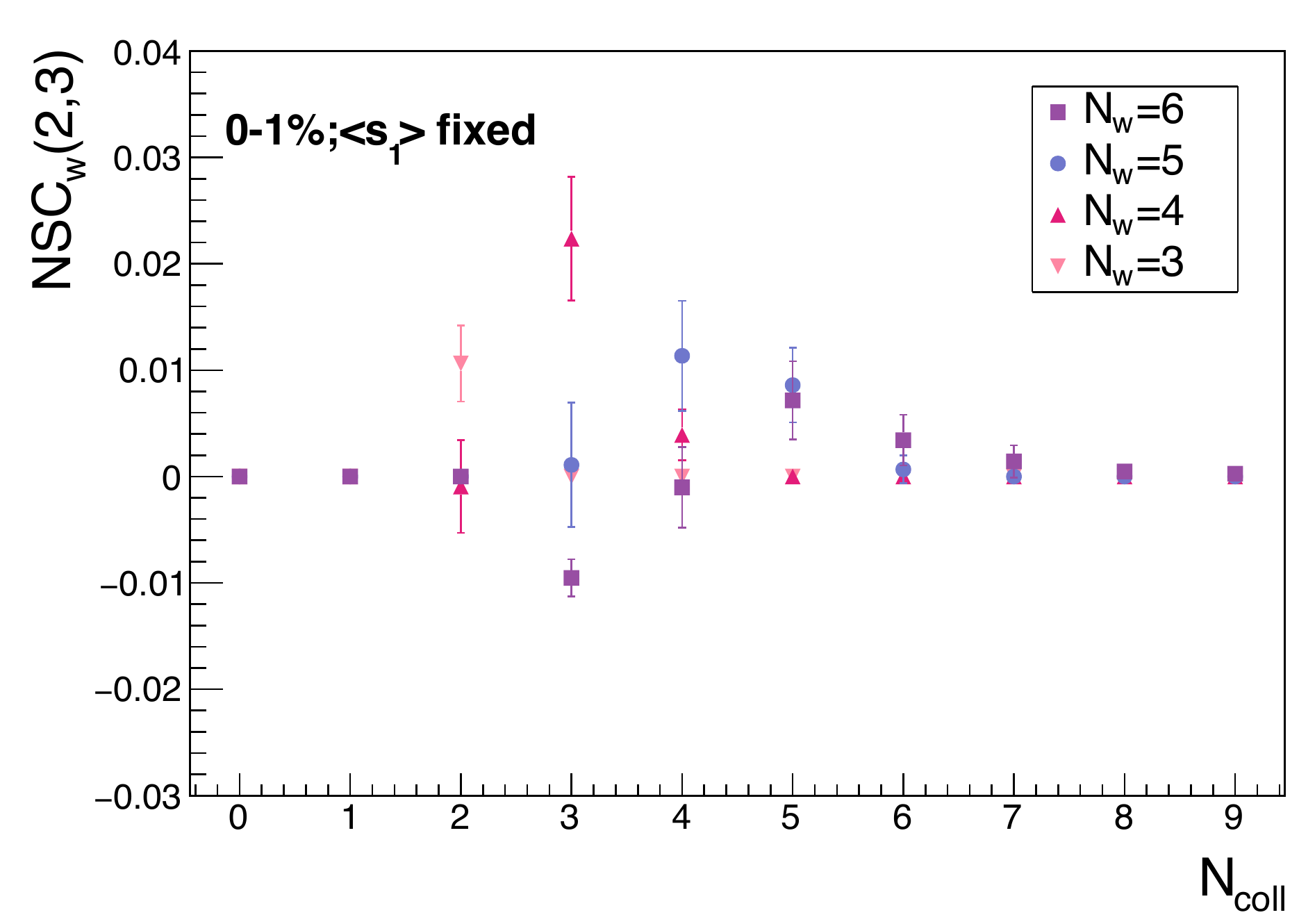} 
\end{center}
\vspace*{-0.5cm}
\caption[a]{Average value of NSC$_{\rm w}$(2,3) as a function of the number of collisions for different number of wounded hot spots. Top: $r_c\!=\!0.4$~fm case. Bottom: $\langle s_1 \rangle$ fixed case.}
\label{nsc23_nw}
\end{figure}

Far from being a casual coincidence or an artifact this effect is observed for any number of wounded hot spots as it is depicted in Fig.~\ref{nsc23_nw}. In the top pannel we show the event-averaged value of NSC$_{\rm w}$(2,3) with respect to the number of collisions for $N_w\!=\!3$ to $N_{w}\!=\!6$ for the $r_c\!=\!0.4$ case. Once again, the configurations that contribute more to the total value of NSC(2,3) are the ones with a large number of wounded hot spots that interact a small amount of times. In opposition, as displayed in the bottom pannel, the interaction topologies that have associated a negative NSC(2,3) are extremely suppressed in the uncorrelated scenario where the configuration with the biggest weight and precisely positive value of NSC(2,3) is ($N_{w}\!=\!4, N_{\rm coll}\!=\!3$).

Then, by computing NSC$_{\rm w}$(2,3) for the different interaction topologies we find that the origin of the negative sign of NSC(2,3) in the $r_c\!=\!0.4$ scenario is due to the decisive role of correlations in modifying the weights of the diverse configurations in the Monte-Carlo Glauber simulations.

\subsection{Scan of the parameter space}
\label{param_space}
To conclude our study we check the sensitivity of the obtained results on the values of the model parameters. Thus, we focus on the correlated scenario and study the dependence of NSC(2,3) on the radius of the hot spot and the repulsive core distance in the [0-1\%] centrality bin. As it could be argued that $r_{c}\!=\!0.4$~fm is a large repulsive distance that may be unrealistic we explore the results of our model for $r_{c}\!=\!0.25$~fm. In the case of $R_{hs}$, we choose 4 different values in our scan $\lbrace 0.15,0.25,0.32,0.4 \rbrace$~fm. Consequently, the parameters of the Gamma distribution for the entropy deposition (see Eq.9 in \cite{Albacete:2016gxu}) are extracted in all the cases by fitting the experimental charged-particle multiplicity distributions $\mathcal P(N_{\rm ch})$ \cite{Aaboud:2016itf}. The other two parameters of our model, namely $R$ and $\rho_{hs}$, remain fixed to their default values given in Table~\ref{param}. Except in the chosen values for $R_{hs}$ and $r_c$ appearing in Table~\ref{param}, i.e. $R_{hs}\!=\!0.32$~fm and $r_c\!=0.4\!$~fm, the requirement that our model reproduces the p+p total cross section and $\rho$ is not fulfilled. Removing these phenomenological constraints allows to pinpoint the effect of just varying the radius of the hot spot or the correlation distance in our results.
\begin{figure}[htb]
\begin{center}
\includegraphics[scale=0.465]{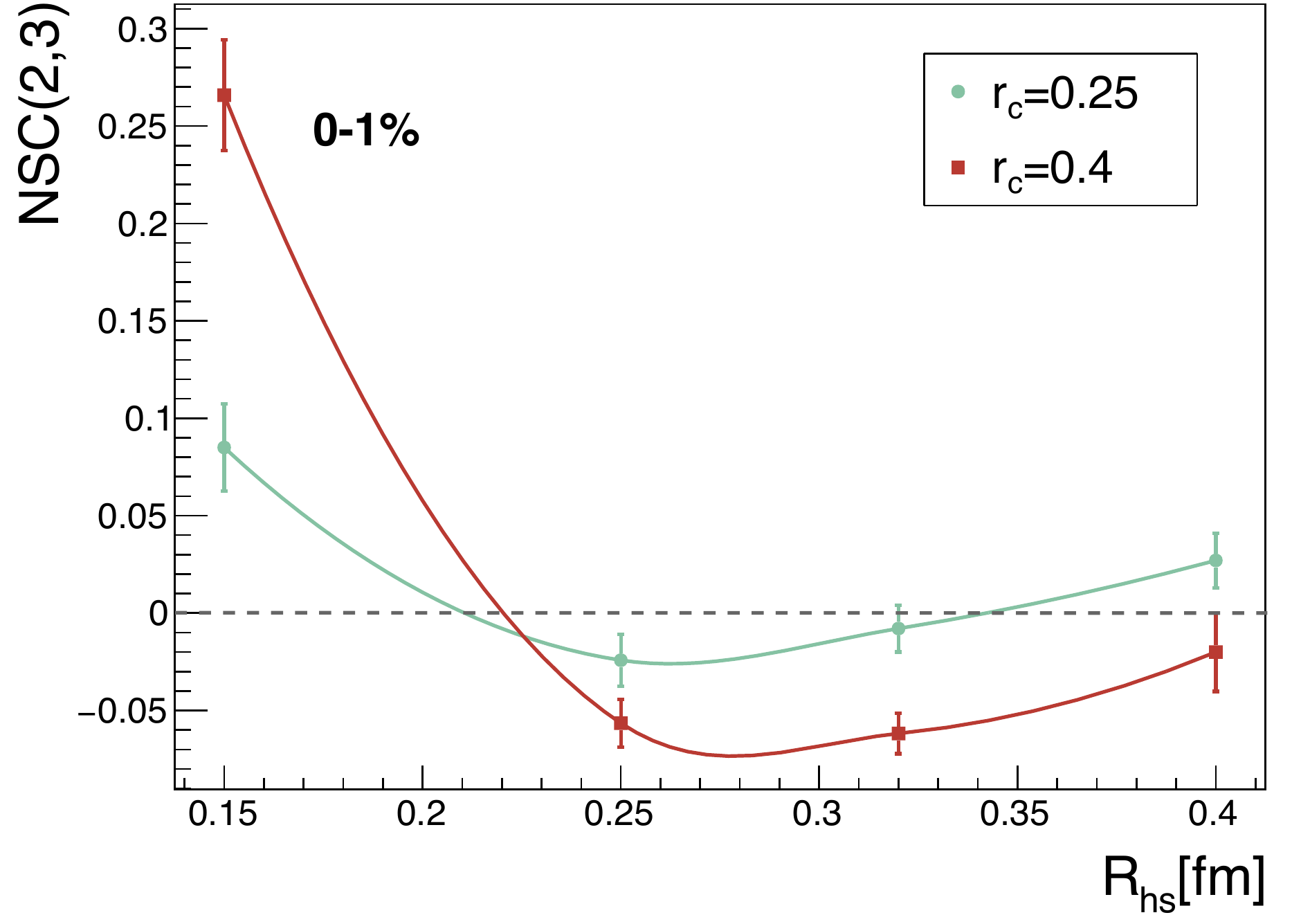} 
\end{center}
\vspace*{-0.5cm}
\caption[a]{Average value of NSC(2,3) as a function of the radius of the hot spot for two different values of the repulsive distance: $r_c\!=\!0.25$~fm (filled green circles) and  $r_c\!=\!0.4$~fm (filled red squares).}
\label{nsc23_rhs}
\end{figure}

In Fig.~\ref{nsc23_rhs} we represent the event-averaged value of NSC(2,3) as a function of $R_{hs}$ for the two different values of the correlation distance considered. First, we observe that by reducing the value of $r_c$ for a given value of $R_{hs}$ we get closer to the uncorrelated case and thus the value of NSC(2,3) is enlarged and pushed to the positive regime, as expected. However, this statement is not universal as it breaks down when $R_{hs}\lesssim 0.22$~fm. In this scenario of very small values of the radius of the hot spot, i.e. $R_{hs}\lesssim 0.22$~fm, $\varepsilon_2$ and $\varepsilon_3$ are positively correlated for both values of the repulsive core distance and the value of NSC(2,3) is larger in the $r_c\!=\!0.4$ case. This result indicates that NSC(2,3) is not sensitive to $R_{hs}$ and $r_c$ independently but to the interplay of both scales. In other words, NSC(2,3) depends on a generic function of the radius of the hot spot and the repulsive core distance $f(R_{hs},r_c)$.  
\begin{figure}[htb]
\begin{center}
\includegraphics[scale=0.465]{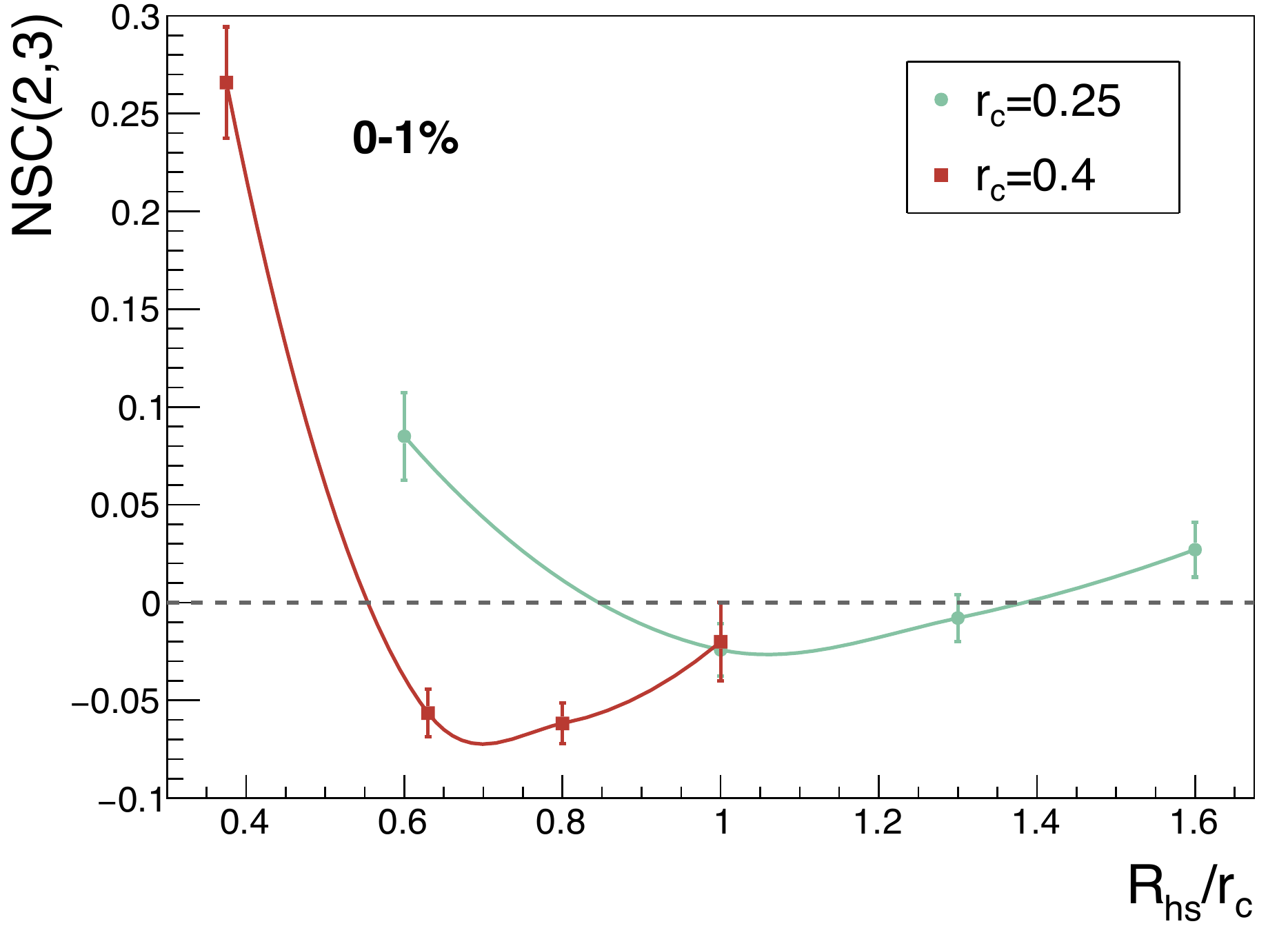} 
\end{center}
\vspace*{-0.5cm}
\caption[a]{Average value of NSC(2,3) as a function of the ratio $R_{hs}/r_c$ for two different values of the repulsive distance: $r_c\!=\!0.25$~fm (filled green circles) and  $r_c\!=\!0.4$~fm (filled red squares).}
\label{nsc23_rhsrc}
\end{figure}

As a first and simple guess to the functional form of $f(R_{hs},r_c)$ we choose it to be the ratio of the two scales involved i.e. $f(R_{hs},r_c)\!=\! R_{hs}/r_c$. This ratio has a transparent interpretation by characterizing the degree of repulsion: if $R_{hs}/r_c\gg 1$ the shape of the proton resembles the uncorrelated scenario where the hot spots can largely overlap in transverse space. The results of NSC(2,3) for different values of $R_{hs}/r_c$ are displayed in Fig.~\ref{nsc23_rhsrc}. We can distinguish three regimes in this plot. On the one hand, when $R_{hs}/r_c\ge 1$ the geometric picture of the proton approaches the uncorrelated scenario and the value of NSC(2,3) increases monotonically starting to be positive for $R_{hs}/r_c\gtrsim 1.3$. Moreover, when $R_{hs}/r_c\!=\!1$ the value of NSC(2,3) is identical within error bars for both correlation scenarios $r_c\!=\!0.25$ and $r_c\!=\!0.4$ supporting the idea that NSC(2,3) depends on $f(R_{hs},r_c)\!=\! R_{hs}/r_c$. However, on the second regime characterized by $0.6\lesssim R_{hs}/r_c \lesssim 1$, we find an abrupt change of the value of NSC(2,3) when slightly increasing the ratio $R_{hs}/r_c$ from 0.6 to 0.63. This suggests that a residual dependence of NSC(2,3) on the other scale of the problem, $R$, may exist. 
Finally, configurations in which the hot spots are much smaller than the repulsive core distance between them i.e. $R_{hs}/r_c\lesssim 0.6$ result into a positive correlation between $\varepsilon_2$ and $\varepsilon_3$. Then, our study favors values of $0.6\lesssim R_{hs}/r_c \lesssim 1.3$ in order to be compatible with the experimental observation of NSC(2,3)$<0$ in the highest centrality bin. Unfortunately, this interval is large enough to be compatible with a picture of the proton in which the hot spots transverse separation is larger than in the uncorrelated case but still they can overlap ($R_{hs}/r_c\sim 1.3$) and with a much more dilute description in which the probability of two hot spots to overlap is highly supressed ($R_{hs}/r_c\sim 0.6$).

\section{Sensitivity of NSC(2,3) to $N_{hs}$}
\label{nhs}

\begin{table}
\begin{center}
\begin{tabular}{c|c|*{5}{c|}} 
&
 \multicolumn{4}{|c|} {\bf Correlated} 
  & \multicolumn{1}{|c|} {$\bf{\langle s_1 \rangle}$ \bf fixed} \\ \toprule[0.5mm]
&  $R_{hs}$~[fm]&$R$~[fm] & $r_c$~[fm] & $ R_{p}$~[fm] & $R$~[fm] \\ \toprule[0.5mm]
 ${\bf N_{hs}\!=\! 2}$& 0.51&1.04 & 0.35 & 1.31 & 1.13  \\ \toprule[0.5mm]
 ${\bf N_{hs}\!=\! 4}$& 0.21&0.55 & 0.32 & 1.2 & 0.64  
\end{tabular}
\caption{Default values of the parameters characterizing the hot spots distribution Eq.~\ref{corr} and their probability to interact Eq.~\ref{gin} for different number of hot spots both in the correlated and "$\langle s_1 \rangle$ fixed" cases.}
\label{param_2}
\end{center}
\end{table}

\begin{table}
\begin{center}
\begin{tabular}{c|c|*{6}{c|}} 
&
 \multicolumn{5}{|c|} {\bf Correlated}  \\ \toprule[0.5mm]
&  $\overline n_1$&$\kappa_1$ & $\overline n_2$  & $\kappa_2$ & $\alpha$  \\ \toprule[0.5mm]
 ${\bf N_{hs}\!=\! 2}$& 26.22 &1.21 & 4.64 & 1.79 & 0.45\\ \toprule[0.5mm]
 ${\bf N_{hs}\!=\! 4}$& 25.68& 1.46 & 4.54 & 1.82 & 0.34\\ \toprule[0.5mm]
\end{tabular}

\begin{tabular}{c|c|*{6}{c|}} 
\toprule[0.5mm]
&
 \multicolumn{5}{|c|} {$\bf{\langle s_1 \rangle}$ \bf fixed }  \\ \toprule[0.5mm]
&  $\overline n_1$&$\kappa_1$ & $\overline n_2$  & $\kappa_2$ & $\alpha$  \\ \toprule[0.5mm]
 ${\bf N_{hs}\!=\! 2}$& 24.66 &1.05 & 4.55 & 2.05 & 0.49\\ \toprule[0.5mm]
 ${\bf N_{hs}\!=\! 4}$& 23.04& 1.15 & 4.37 & 2.14 & 0.4
\end{tabular}
\caption{Default values of the parameters of the double Gamma distribution that characterizes the fluctuating amount of entropy each wounded hot spot deposits, $s_0$, for different number of hot spots both in the correlated (top) and "$\langle s_1 \rangle$ fixed" cases (bottom).}
\label{param_2_entropy}
\end{center}
\end{table}

\begin{figure}[htb]
\begin{center}
\includegraphics[scale=0.465]{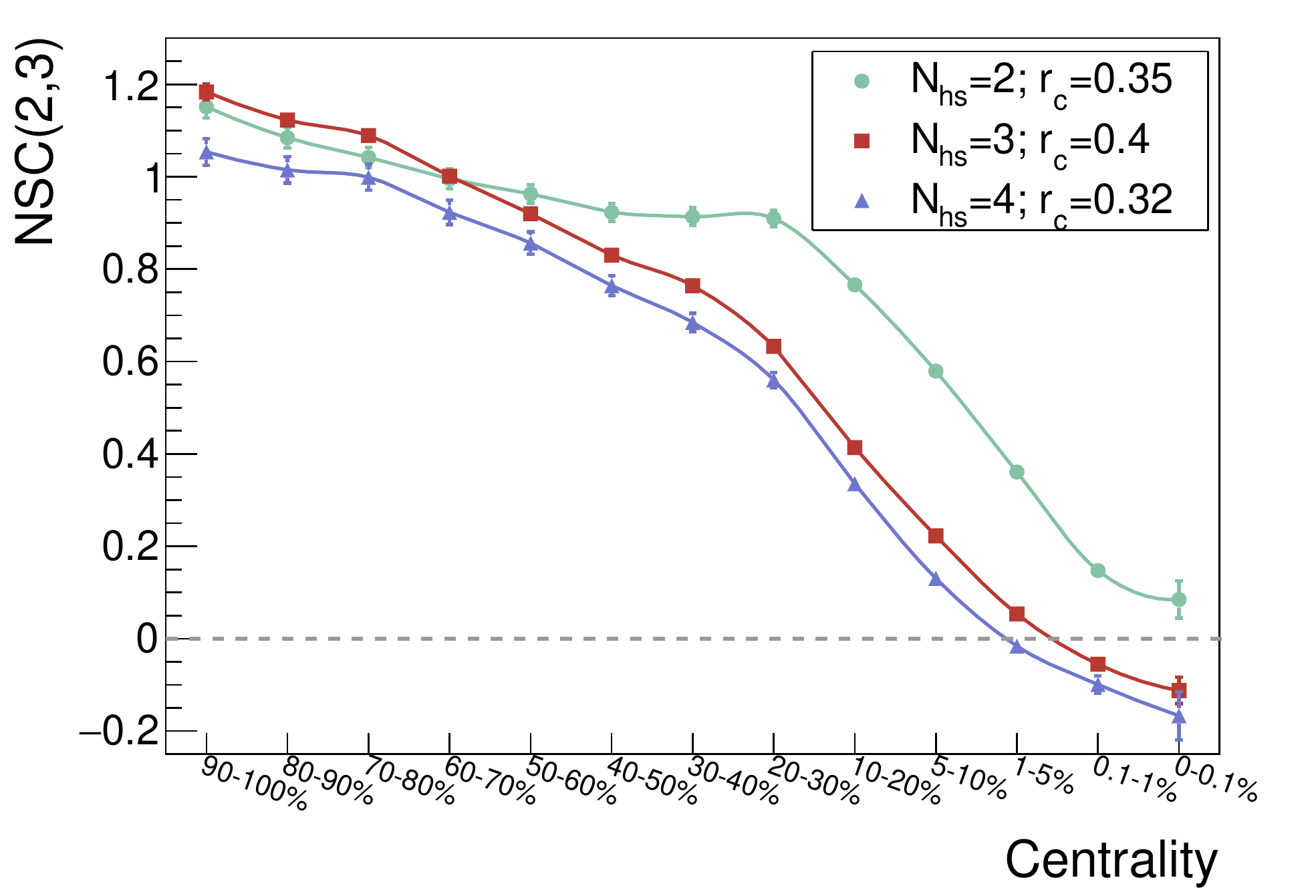} 
\end{center}
\vspace*{-0.5cm}
\caption[a]{Average value of NSC(2,3) as a function of the centrality range in the correlated scenario for $N_{hs}\!=\!2$ (filled green circles), $N_{hs}\!=\!3$ and (filled red squares) $N_{hs}\!=\!4$ (filled violet triangles). The error bars represent statistical uncertainties.}
\label{nsc23_nhs}
\end{figure}

All along the manuscript we have considered that the proton is constituted by 3 gluonic hot spots. This is the most natural scenario when a direct correspondance between the Fock space of valence partons and the hot spots is assumed. However, this relation is arguable as, while being extensively used as a phenomenological tool, the ultimate dynamical origin of the hot spots remains as an open debate. Therefore, it is opportune to check the reliability of our results after variations of this parameter, $N_{hs}$. 

Following the ideas of the previous sections, we focus our discussion on the results for NSC(2,3) after considering the two more straight-forward extensions of our model: $N_{hs}\!=\!2$ and $N_{hs}\!=\!4$. In order to make a fair comparison between the three different scenarios, $N_{hs}\!=\!(2,3,4)$, we choose representative values of the parameters $\lbrace{ R_{hs},R \rbrace }$ that fulfill two constraints. As in the previous sections, the experimental value of the total p+p cross section is reproduced. Further, the proton radius defined as  
$R_p\!=\!\sqrt{N_{hs}}\sqrt{\langle s_1^2 \rangle+R_{hs}^2}$, where $\langle s_1 \rangle$ is the r.m.s of the spatial probability distribution given by Eq.~\ref{corr}, should not depend on the number of hot spots that the proton contains so we fix it to be the same in all the cases. All in all, the values of the parameters for the correlated and $\langle s_1 \rangle$ fixed cases are given in Tables~\ref{param_2} and \ref{param_2_entropy}.

We start by exploring the dependence of the event-averaged value of NSC(2,3) on the number of hot spots in the correlated scenarios as displayed in Fig.~\ref{nsc23_nhs} as a function of centrality. The differences between the three cases start to appear in mid-to-ultra central collisions. There exists a clear trend towards smaller values of NSC(2,3) when a bigger number of hot spots is considered. Specifically, the negative sign of NSC(2,3) in the high centrality bins is not achieved when $N_{hs}\!=\!2$ even with correlations. Thus, we conclude that with the selected parameters the minimum number of hot spots to describe the onset of the anti-correlation between $\varepsilon_2$ and $\varepsilon_3$ is $N_{hs}\!=\!3$. It should also be noted that the inclusion of an additional hot spot i.e. $N_{hs}\!=\!4$ helps to make NSC(2,3) even more negative in the highest centrality bins although the effect is small when compared to the drastic impact of changing from $N_{hs}\!=\!2$ to $N_{hs}\!=\!3$. 
\begin{figure}[htb]
\begin{center}
\includegraphics[scale=0.465]{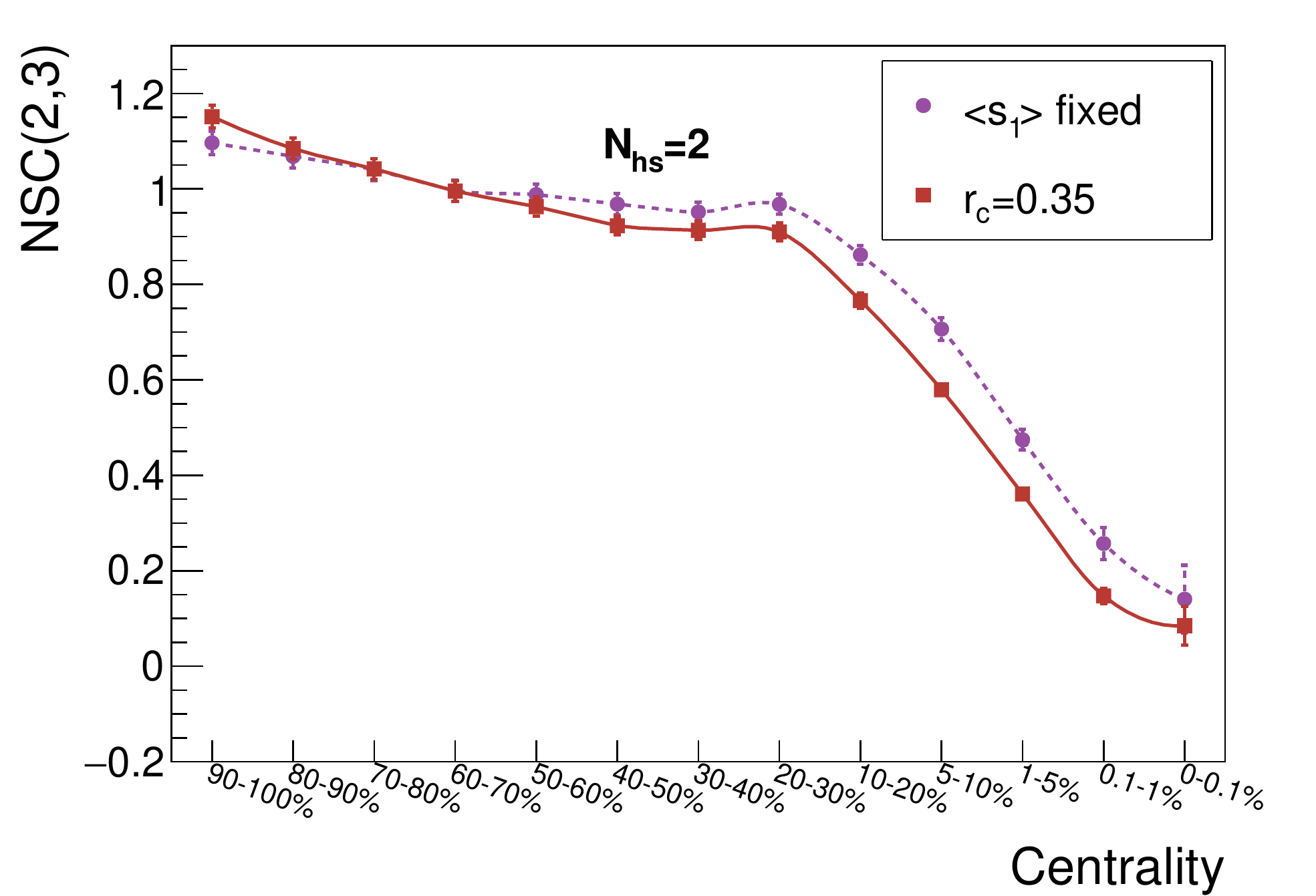} 
\includegraphics[scale=0.465]{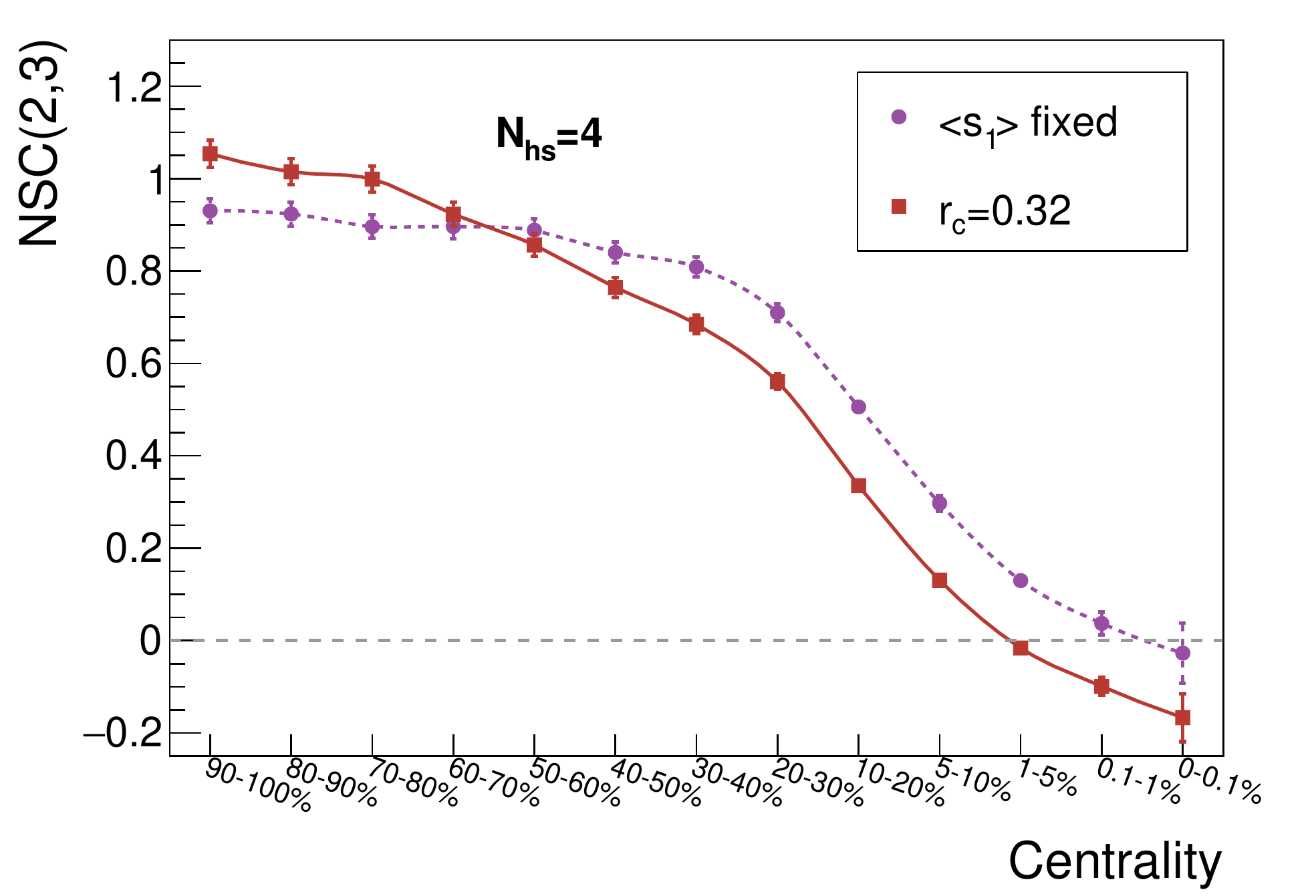}
\end{center}
\vspace*{-0.5cm}
\caption[a]{Average value of NSC(2,3) as a function of the centrality range for $\langle s_1 \rangle$ fixed (purple short-dashed line connecting filled purple circles) and $r_c\!=\!0.4$~fm (red solid line connecting filled red squares). The error bars represent statistical uncertainties. Top: $N_{hs}\!=\!2$. Bottom: $N_{hs}\!=\!4$. }
\label{nsc23_nhs2}
\end{figure}

To conclude the discussion on the sensitivity of our model to the number of gluonic hot spots, in Fig.~\ref{nsc23_nhs2} the comparison between the correlated and $\langle s_1 \rangle$ fixed scenarios for $N_{hs}\!=\!2$ (top) and $N_{hs}\!=\!4$ (bottom) is displayed. First and foremost, the effect of including spatial correlations is invariant under changes in the number of hot spots: in peripheral collisions they enlarge the positive correlation of $\varepsilon_2$ and $\varepsilon_3$ while favoring a negative sign of NSC(2,3) with respect to the uncorrelated scenario. Albeit the correlated curve is always below the uncorrelated scenario in the highest centrality bins an important comment is in order: NSC(2,3) is compatible with negative values, within statistical uncertainty, in the [0-0.1\%] bin for the uncorrelated case. This fact reinforce the idea remarked in Sec.~\ref{param_space}: the interplay of the different scales $\lbrace { R_{hs},r_c,N_{hs}\rbrace }$ is decisive in the sign of NSC(2,3) within our framework. For $N_{hs}\!=\!4$ the weight of the configurations with a large number of wounded hot spots and a small number of collisions is large enough so that the spatial correlations are not essential to obtain a negative NSC(2,3) in the $[0\!-\!0.1\%]$ bin. However, as these configurations are enhanced in the correlated scenario, the anti-correlation of $\varepsilon_2$ and $\varepsilon_3$ is stronger than in the uncorrelated case just as in the case of $N_{hs}\!=\!3$. Finally, we have checked that the effect of spatial correlations on the average values of $\varepsilon_2$ and $\varepsilon_3$ is qualitatively the same as in the $N_{hs}\!=\!3$ case studied in \cite{Albacete:2016gxu}. To sum up, although the negative sign of NSC(2,3) in the highest centrality bins is not a unique feature of the correlated scenario but relies on the interplay of the different scales, the inclusion of repulsive correlations provides a mechanism to reduce its value in the highest centrality bins. 

\section{Discussion and Outlook}
\label{final}
The experimental finding of a resembling behavior of the fluctuations of the Fourier harmonic coefficients $v_n$, in terms of the symmetric cumulants SC(n,m) in p+p, p+Pb and Pb+Pb collisions at LHC energies constitutes a new piece of the puzzle on whether collective effects are being observed in small systems. Data suggests that the fluctuations of $v_2$ and $v_3$ are anti-correlated i.e. SC(2,3)$<0$ in proton-proton collisions with $N_{\rm trk}^{\rm offline}\sim 100$. In this very-high multiplicity regime, where the non-flow contributions arising from jet correlations are subleading, the value of the symmetric cumulant is sensitive to the initial state fluctuations of the collision.

In this article, we perform a systematic study of the normalized symmetric cumulants NSC(2,3) and NSC(2,4) in proton-proton interactions at $\sqrt s\!=\! 13$~TeV in terms of the initial geometry of the collision. We rely on the wounded hot spot model used in \cite{Albacete:2016gxu} where the proton is regarded as a system of three gluonic hot spots whose transverse positions are not independent but correlated via a repulsive core distance. We find that the effect of these repulsive correlations is key, within our relatively simple geometric model, to obtain an anti-correlation of the fluctuations of $\varepsilon_2$ and $\varepsilon_3$ in the highest centrality bin. More precisely, they enhance the probability of having interactions with a large number of wounded hot spots colliding a small amount of times. These interaction topologies are responsible in our set up of the negative sign of the symmetric cumulant NSC(2,3) in the correlated scenario. Further we explore the dependence of NSC(2,3) on the values of the repulsive distance and the radius of the hot spot concluding that it is not a function of these two variables independently but to a combination of both such as their ratio $R_{hs}/r_c$. We also show that the values of NSC(2,3) are sensitive to the variation of the number of hot spots that constitute the proton. Specifically, we find that, within the correlated scenario, adding an extra hot spot to our description reinforces the negative sign of NSC(2,3) while reducing it to $N_{hs}\!=\!2$ pushes NSC(2,3) towards positive values. Moreover, the enhanced probability of having configurations with a large $N_w/N_{\rm coll}$ when $N_{hs}\!=\!4$ permits a negative value of NSC(2,3) within the uncorrelated scenario although its absolute value is smaller than in the correlated case. This fact reinforces the argument that the sign of NSC(2,3) is sensitive to the interplay of the different scales. 

Our study confirms the idea that NSC(2,3) in proton-proton interactions is extremely sensitive to the initial state fluctuations. Further, as we have shown, it can help to discriminate between different parameterizations of the proton's geometry and, concretely, to realize the importance of spatial correlations. A precise characterization of the proton's geometry has a direct impact in the flow studies in proton-proton collisions. Furthermore, parametrizing how are the subnucleonic degrees of freedom arranged inside the proton is an essential input in event generators that aim to describe multi-parton interactions, the mechanism that dominates the underlying event at LHC energies.

In order to confirm the conclusions exposed in this manuscript the natural continuation of this work would consist on feeding a relativistic viscous hydrodynamic simulation with our initial entropy density profiles and check if the effect of these spatial correlations is washed out by the evolution or, on the contrary, it impacts the values of the Fourier flow coefficients.  \\

\section*{ACKNOWLEDGMENTS} 

AS would like to thank Maxime Guilbaud and Antonio Bueno for profitable and enlightening discussions. Further, AS thanks the Dpto. de F\'isica Te\'orica y del Cosmos in the University of Granada for the warming hospitality during the early stages of this work thanks to a grant from the COST Action CA15213 THOR. This work was partially supported by a Helmholtz Young Investigator Group VH-NG-822 from the Helmholtz Association and GSI, the Helmholtz International
Center for the Facility for Antiproton and Ion Research (HIC for FAIR) within the
framework of the Landes-Offensive zur Entwicklung Wissenschaftlich-Oekonomischer Exzellenz (LOEWE) program launched by the State of Hesse, a FP7-PEOPLE-2013-CIG Grant of the European Commission, reference QCDense/631558, and by Ram\'on y Cajal and MINECO projects reference RYC-2011-09010 and FPA2013-47836. 

\bibliography{ref_cumulants}{}

\begin{thebibliography}{43}%
\makeatletter
\providecommand \@ifxundefined [1]{%
 \@ifx{#1\undefined}
}%
\providecommand \@ifnum [1]{%
 \ifnum #1\expandafter \@firstoftwo
 \else \expandafter \@secondoftwo
 \fi
}%
\providecommand \@ifx [1]{%
 \ifx #1\expandafter \@firstoftwo
 \else \expandafter \@secondoftwo
 \fi
}%
\providecommand \natexlab [1]{#1}%
\providecommand \enquote  [1]{``#1''}%
\providecommand \bibnamefont  [1]{#1}%
\providecommand \bibfnamefont [1]{#1}%
\providecommand \citenamefont [1]{#1}%
\providecommand \href@noop [0]{\@secondoftwo}%
\providecommand \href [0]{\begingroup \@sanitize@url \@href}%
\providecommand \@href[1]{\@@startlink{#1}\@@href}%
\providecommand \@@href[1]{\endgroup#1\@@endlink}%
\providecommand \@sanitize@url [0]{\catcode `\\12\catcode `\$12\catcode
  `\&12\catcode `\#12\catcode `\^12\catcode `\_12\catcode `\%12\relax}%
\providecommand \@@startlink[1]{}%
\providecommand \@@endlink[0]{}%
\providecommand \url  [0]{\begingroup\@sanitize@url \@url }%
\providecommand \@url [1]{\endgroup\@href {#1}{\urlprefix }}%
\providecommand \urlprefix  [0]{URL }%
\providecommand \Eprint [0]{\href }%
\providecommand \doibase [0]{http://dx.doi.org/}%
\providecommand \selectlanguage [0]{\@gobble}%
\providecommand \bibinfo  [0]{\@secondoftwo}%
\providecommand \bibfield  [0]{\@secondoftwo}%
\providecommand \translation [1]{[#1]}%
\providecommand \BibitemOpen [0]{}%
\providecommand \bibitemStop [0]{}%
\providecommand \bibitemNoStop [0]{.\EOS\space}%
\providecommand \EOS [0]{\spacefactor3000\relax}%
\providecommand \BibitemShut  [1]{\csname bibitem#1\endcsname}%
\let\auto@bib@innerbib\@empty
\bibitem [{\citenamefont {Albacete}\ \emph {et~al.}(2017)\citenamefont
  {Albacete}, \citenamefont {Petersen},\ and\ \citenamefont
  {Soto-Ontoso}}]{Albacete:2016gxu}%
  \BibitemOpen
  \bibfield  {author} {\bibinfo {author} {\bibfnamefont {J.~L.}\ \bibnamefont
  {Albacete}}, \bibinfo {author} {\bibfnamefont {H.}~\bibnamefont {Petersen}},
  \ and\ \bibinfo {author} {\bibfnamefont {A.}~\bibnamefont {Soto-Ontoso}},\
  }\href {\doibase 10.1103/PhysRevC.95.064909} {\bibfield  {journal} {\bibinfo
  {journal} {Phys. Rev.}\ }\textbf {\bibinfo {volume} {C95}},\ \bibinfo {pages}
  {064909} (\bibinfo {year} {2017})},\ \Eprint
  {http://arxiv.org/abs/1612.06274} {arXiv:1612.06274 [hep-ph]} \BibitemShut
  {NoStop}%
\bibitem [{\citenamefont {Bilandzic}\ \emph {et~al.}(2014)\citenamefont
  {Bilandzic}, \citenamefont {Christensen}, \citenamefont {Gulbrandsen},
  \citenamefont {Hansen},\ and\ \citenamefont {Zhou}}]{Bilandzic:2013kga}%
  \BibitemOpen
  \bibfield  {author} {\bibinfo {author} {\bibfnamefont {A.}~\bibnamefont
  {Bilandzic}}, \bibinfo {author} {\bibfnamefont {C.~H.}\ \bibnamefont
  {Christensen}}, \bibinfo {author} {\bibfnamefont {K.}~\bibnamefont
  {Gulbrandsen}}, \bibinfo {author} {\bibfnamefont {A.}~\bibnamefont {Hansen}},
  \ and\ \bibinfo {author} {\bibfnamefont {Y.}~\bibnamefont {Zhou}},\ }\href
  {\doibase 10.1103/PhysRevC.89.064904} {\bibfield  {journal} {\bibinfo
  {journal} {Phys. Rev.}\ }\textbf {\bibinfo {volume} {C89}},\ \bibinfo {pages}
  {064904} (\bibinfo {year} {2014})},\ \Eprint {http://arxiv.org/abs/1312.3572}
  {arXiv:1312.3572 [nucl-ex]} \BibitemShut {NoStop}%
\bibitem [{\citenamefont {Di~Francesco}\ \emph {et~al.}(2017)\citenamefont
  {Di~Francesco}, \citenamefont {Guilbaud}, \citenamefont {Luzum},\ and\
  \citenamefont {Ollitrault}}]{DiFrancesco:2016srj}%
  \BibitemOpen
  \bibfield  {author} {\bibinfo {author} {\bibfnamefont {P.}~\bibnamefont
  {Di~Francesco}}, \bibinfo {author} {\bibfnamefont {M.}~\bibnamefont
  {Guilbaud}}, \bibinfo {author} {\bibfnamefont {M.}~\bibnamefont {Luzum}}, \
  and\ \bibinfo {author} {\bibfnamefont {J.-Y.}\ \bibnamefont {Ollitrault}},\
  }\href {\doibase 10.1103/PhysRevC.95.044911} {\bibfield  {journal} {\bibinfo
  {journal} {Phys. Rev.}\ }\textbf {\bibinfo {volume} {C95}},\ \bibinfo {pages}
  {044911} (\bibinfo {year} {2017})},\ \Eprint
  {http://arxiv.org/abs/1612.05634} {arXiv:1612.05634 [nucl-th]} \BibitemShut
  {NoStop}%
\bibitem [{\citenamefont {Giacalone}\ \emph {et~al.}(2016)\citenamefont
  {Giacalone}, \citenamefont {Yan}, \citenamefont {Noronha-Hostler},\ and\
  \citenamefont {Ollitrault}}]{Giacalone:2016afq}%
  \BibitemOpen
  \bibfield  {author} {\bibinfo {author} {\bibfnamefont {G.}~\bibnamefont
  {Giacalone}}, \bibinfo {author} {\bibfnamefont {L.}~\bibnamefont {Yan}},
  \bibinfo {author} {\bibfnamefont {J.}~\bibnamefont {Noronha-Hostler}}, \ and\
  \bibinfo {author} {\bibfnamefont {J.-Y.}\ \bibnamefont {Ollitrault}},\ }\href
  {\doibase 10.1103/PhysRevC.94.014906} {\bibfield  {journal} {\bibinfo
  {journal} {Phys. Rev.}\ }\textbf {\bibinfo {volume} {C94}},\ \bibinfo {pages}
  {014906} (\bibinfo {year} {2016})},\ \Eprint
  {http://arxiv.org/abs/1605.08303} {arXiv:1605.08303 [nucl-th]} \BibitemShut
  {NoStop}%
\bibitem [{\citenamefont {Adam}\ \emph {et~al.}(2016)\citenamefont {Adam} \emph
  {et~al.}}]{ALICE:2016kpq}%
  \BibitemOpen
  \bibfield  {author} {\bibinfo {author} {\bibfnamefont {J.}~\bibnamefont
  {Adam}} \emph {et~al.} (\bibinfo {collaboration} {ALICE}),\ }\href {\doibase
  10.1103/PhysRevLett.117.182301} {\bibfield  {journal} {\bibinfo  {journal}
  {Phys. Rev. Lett.}\ }\textbf {\bibinfo {volume} {117}},\ \bibinfo {pages}
  {182301} (\bibinfo {year} {2016})},\ \Eprint
  {http://arxiv.org/abs/1604.07663} {arXiv:1604.07663 [nucl-ex]} \BibitemShut
  {NoStop}%
\bibitem [{\citenamefont {Collaboration}(2017)}]{CMS:2017saf}%
  \BibitemOpen
  \bibfield  {author} {\bibinfo {author} {\bibfnamefont {C.}~\bibnamefont
  {Collaboration}} (\bibinfo {collaboration} {CMS}),\ }\href@noop {} {\
  (\bibinfo {year} {2017})}\BibitemShut {NoStop}%
\bibitem [{\citenamefont {Khachatryan}\ \emph {et~al.}(2017)\citenamefont
  {Khachatryan} \emph {et~al.}}]{Khachatryan:2016txc}%
  \BibitemOpen
  \bibfield  {author} {\bibinfo {author} {\bibfnamefont {V.}~\bibnamefont
  {Khachatryan}} \emph {et~al.} (\bibinfo {collaboration} {CMS}),\ }\href
  {\doibase 10.1016/j.physletb.2016.12.009} {\bibfield  {journal} {\bibinfo
  {journal} {Phys. Lett.}\ }\textbf {\bibinfo {volume} {B765}},\ \bibinfo
  {pages} {193} (\bibinfo {year} {2017})}\BibitemShut {NoStop}%
\bibitem [{\citenamefont {Aad}\ \emph {et~al.}(2016)\citenamefont {Aad} \emph
  {et~al.}}]{Aad:2015gqa}%
  \BibitemOpen
  \bibfield  {author} {\bibinfo {author} {\bibfnamefont {G.}~\bibnamefont
  {Aad}} \emph {et~al.} (\bibinfo {collaboration} {ATLAS}),\ }\href {\doibase
  10.1103/PhysRevLett.116.172301} {\bibfield  {journal} {\bibinfo  {journal}
  {Phys. Rev. Lett.}\ }\textbf {\bibinfo {volume} {116}},\ \bibinfo {pages}
  {172301} (\bibinfo {year} {2016})}\BibitemShut {NoStop}%
\bibitem [{\citenamefont {Khachatryan}\ \emph {et~al.}(2010)\citenamefont
  {Khachatryan} \emph {et~al.}}]{Khachatryan:2010gv}%
  \BibitemOpen
  \bibfield  {author} {\bibinfo {author} {\bibfnamefont {V.}~\bibnamefont
  {Khachatryan}} \emph {et~al.} (\bibinfo {collaboration} {CMS}),\ }\href
  {\doibase 10.1007/JHEP09(2010)091} {\bibfield  {journal} {\bibinfo  {journal}
  {JHEP}\ }\textbf {\bibinfo {volume} {09}},\ \bibinfo {pages} {091} (\bibinfo
  {year} {2010})}\BibitemShut {NoStop}%
\bibitem [{\citenamefont {Adam}\ \emph {et~al.}(2017)\citenamefont {Adam} \emph
  {et~al.}}]{ALICE:2017jyt}%
  \BibitemOpen
  \bibfield  {author} {\bibinfo {author} {\bibfnamefont {J.}~\bibnamefont
  {Adam}} \emph {et~al.} (\bibinfo {collaboration} {ALICE}),\ }\href {\doibase
  10.1038/nphys4111} {\bibfield  {journal} {\bibinfo  {journal} {Nature Phys.}\
  }\textbf {\bibinfo {volume} {13}},\ \bibinfo {pages} {535} (\bibinfo {year}
  {2017})},\ \Eprint {http://arxiv.org/abs/1606.07424} {arXiv:1606.07424
  [nucl-ex]} \BibitemShut {NoStop}%
\bibitem [{\citenamefont {Schlichting}\ and\ \citenamefont
  {Tribedy}(2016)}]{Schlichting:2016sqo}%
  \BibitemOpen
  \bibfield  {author} {\bibinfo {author} {\bibfnamefont {S.}~\bibnamefont
  {Schlichting}}\ and\ \bibinfo {author} {\bibfnamefont {P.}~\bibnamefont
  {Tribedy}},\ }\href {\doibase 10.1155/2016/8460349} {\bibfield  {journal}
  {\bibinfo  {journal} {Adv. High Energy Phys.}\ }\textbf {\bibinfo {volume}
  {2016}},\ \bibinfo {pages} {8460349} (\bibinfo {year} {2016})},\ \Eprint
  {http://arxiv.org/abs/1611.00329} {arXiv:1611.00329 [hep-ph]} \BibitemShut
  {NoStop}%
\bibitem [{\citenamefont {Weller}\ and\ \citenamefont
  {Romatschke}(2017)}]{Weller:2017tsr}%
  \BibitemOpen
  \bibfield  {author} {\bibinfo {author} {\bibfnamefont {R.~D.}\ \bibnamefont
  {Weller}}\ and\ \bibinfo {author} {\bibfnamefont {P.}~\bibnamefont
  {Romatschke}},\ }\href@noop {} {\  (\bibinfo {year} {2017})},\ \Eprint
  {http://arxiv.org/abs/1701.07145} {arXiv:1701.07145 [nucl-th]} \BibitemShut
  {NoStop}%
\bibitem [{\citenamefont {d'Enterria}\ \emph {et~al.}(2010)\citenamefont
  {d'Enterria}, \citenamefont {Eyyubova}, \citenamefont {Korotkikh},
  \citenamefont {Lokhtin}, \citenamefont {Petrushanko}, \citenamefont
  {Sarycheva},\ and\ \citenamefont {Snigirev}}]{dEnterria:2010xip}%
  \BibitemOpen
  \bibfield  {author} {\bibinfo {author} {\bibfnamefont {D.}~\bibnamefont
  {d'Enterria}}, \bibinfo {author} {\bibfnamefont {G.~K.}\ \bibnamefont
  {Eyyubova}}, \bibinfo {author} {\bibfnamefont {V.~L.}\ \bibnamefont
  {Korotkikh}}, \bibinfo {author} {\bibfnamefont {I.~P.}\ \bibnamefont
  {Lokhtin}}, \bibinfo {author} {\bibfnamefont {S.~V.}\ \bibnamefont
  {Petrushanko}}, \bibinfo {author} {\bibfnamefont {L.~I.}\ \bibnamefont
  {Sarycheva}}, \ and\ \bibinfo {author} {\bibfnamefont {A.~M.}\ \bibnamefont
  {Snigirev}},\ }\href {\doibase 10.1140/epjc/s10052-009-1232-7} {\bibfield
  {journal} {\bibinfo  {journal} {Eur. Phys. J.}\ }\textbf {\bibinfo {volume}
  {C66}},\ \bibinfo {pages} {173} (\bibinfo {year} {2010})},\ \Eprint
  {http://arxiv.org/abs/0910.3029} {arXiv:0910.3029 [hep-ph]} \BibitemShut
  {NoStop}%
\bibitem [{\citenamefont {Dusling}\ \emph {et~al.}(2016)\citenamefont
  {Dusling}, \citenamefont {Li},\ and\ \citenamefont
  {Schenke}}]{Dusling:2015gta}%
  \BibitemOpen
  \bibfield  {author} {\bibinfo {author} {\bibfnamefont {K.}~\bibnamefont
  {Dusling}}, \bibinfo {author} {\bibfnamefont {W.}~\bibnamefont {Li}}, \ and\
  \bibinfo {author} {\bibfnamefont {B.}~\bibnamefont {Schenke}},\ }\href
  {\doibase 10.1142/S0218301316300022} {\bibfield  {journal} {\bibinfo
  {journal} {Int. J. Mod. Phys.}\ }\textbf {\bibinfo {volume} {E25}},\ \bibinfo
  {pages} {1630002} (\bibinfo {year} {2016})},\ \Eprint
  {http://arxiv.org/abs/1509.07939} {arXiv:1509.07939 [nucl-ex]} \BibitemShut
  {NoStop}%
\bibitem [{\citenamefont {Eskola}\ \emph {et~al.}(2017)\citenamefont {Eskola},
  \citenamefont {Niemi}, \citenamefont {Paatelainen},\ and\ \citenamefont
  {Tuominen}}]{Eskola:2017imo}%
  \BibitemOpen
  \bibfield  {author} {\bibinfo {author} {\bibfnamefont {K.~J.}\ \bibnamefont
  {Eskola}}, \bibinfo {author} {\bibfnamefont {H.}~\bibnamefont {Niemi}},
  \bibinfo {author} {\bibfnamefont {R.}~\bibnamefont {Paatelainen}}, \ and\
  \bibinfo {author} {\bibfnamefont {K.}~\bibnamefont {Tuominen}},\ }in\ \href
  {https://inspirehep.net/record/1591543/files/arXiv:1704.04060.pdf} {\emph
  {\bibinfo {booktitle} {{26th International Conference on Ultrarelativistic
  Nucleus-Nucleus Collisions (Quark Matter 2017) Chicago,Illinois, USA,
  February 6-11, 2017}}}}\ (\bibinfo {year} {2017})\ \Eprint
  {http://arxiv.org/abs/1704.04060} {arXiv:1704.04060 [hep-ph]} \BibitemShut
  {NoStop}%
\bibitem [{\citenamefont {Gardim}\ \emph {et~al.}(2017)\citenamefont {Gardim},
  \citenamefont {Grassi}, \citenamefont {Luzum},\ and\ \citenamefont
  {Noronha-Hostler}}]{Gardim:2016nrr}%
  \BibitemOpen
  \bibfield  {author} {\bibinfo {author} {\bibfnamefont {F.~G.}\ \bibnamefont
  {Gardim}}, \bibinfo {author} {\bibfnamefont {F.}~\bibnamefont {Grassi}},
  \bibinfo {author} {\bibfnamefont {M.}~\bibnamefont {Luzum}}, \ and\ \bibinfo
  {author} {\bibfnamefont {J.}~\bibnamefont {Noronha-Hostler}},\ }\href
  {\doibase 10.1103/PhysRevC.95.034901} {\bibfield  {journal} {\bibinfo
  {journal} {Phys. Rev.}\ }\textbf {\bibinfo {volume} {C95}},\ \bibinfo {pages}
  {034901} (\bibinfo {year} {2017})},\ \Eprint
  {http://arxiv.org/abs/1608.02982} {arXiv:1608.02982 [nucl-th]} \BibitemShut
  {NoStop}%
\bibitem [{\citenamefont {Niemi}\ \emph {et~al.}(2016)\citenamefont {Niemi},
  \citenamefont {Eskola},\ and\ \citenamefont {Paatelainen}}]{Niemi:2015qia}%
  \BibitemOpen
  \bibfield  {author} {\bibinfo {author} {\bibfnamefont {H.}~\bibnamefont
  {Niemi}}, \bibinfo {author} {\bibfnamefont {K.~J.}\ \bibnamefont {Eskola}}, \
  and\ \bibinfo {author} {\bibfnamefont {R.}~\bibnamefont {Paatelainen}},\
  }\href {\doibase 10.1103/PhysRevC.93.024907} {\bibfield  {journal} {\bibinfo
  {journal} {Phys. Rev.}\ }\textbf {\bibinfo {volume} {C93}},\ \bibinfo {pages}
  {024907} (\bibinfo {year} {2016})},\ \Eprint
  {http://arxiv.org/abs/1505.02677} {arXiv:1505.02677 [hep-ph]} \BibitemShut
  {NoStop}%
\bibitem [{\citenamefont {Broniowski}\ \emph {et~al.}(2016)\citenamefont
  {Broniowski}, \citenamefont {Bo{\.z}zek},\ and\ \citenamefont
  {Rybczy{\'n}ski}}]{Broniowski:2016pvx}%
  \BibitemOpen
  \bibfield  {author} {\bibinfo {author} {\bibfnamefont {W.}~\bibnamefont
  {Broniowski}}, \bibinfo {author} {\bibfnamefont {P.}~\bibnamefont
  {Bo{\.z}zek}}, \ and\ \bibinfo {author} {\bibfnamefont {M.}~\bibnamefont
  {Rybczy{\'n}ski}},\ }in\ \href
  {https://inspirehep.net/record/1495465/files/arXiv:1611.00250.pdf} {\emph
  {\bibinfo {booktitle} {{10th International Workshop on Critical Point and
  Onset of Deconfinement (CPOD 2016) Wroclaw, Poland, May 30-June 4, 2016}}}}\
  (\bibinfo {year} {2016})\ \Eprint {http://arxiv.org/abs/1611.00250}
  {arXiv:1611.00250 [nucl-th]} \BibitemShut {NoStop}%
\bibitem [{\citenamefont {M.Guilbaud}()}]{talk2}%
  \BibitemOpen
  \bibfield  {author} {\bibinfo {author} {\bibnamefont {M.Guilbaud}},\
  }\href@noop {} {\emph {\bibinfo {title} {Talk presented at Quark Matter
  2017,Chicago (USA),February 05-11, 2017}}}\BibitemShut {NoStop}%
\bibitem [{\citenamefont {Welsh}\ \emph {et~al.}(2016)\citenamefont {Welsh},
  \citenamefont {Singer},\ and\ \citenamefont {Heinz}}]{Welsh:2016siu}%
  \BibitemOpen
  \bibfield  {author} {\bibinfo {author} {\bibfnamefont {K.}~\bibnamefont
  {Welsh}}, \bibinfo {author} {\bibfnamefont {J.}~\bibnamefont {Singer}}, \
  and\ \bibinfo {author} {\bibfnamefont {U.~W.}\ \bibnamefont {Heinz}},\ }\href
  {\doibase 10.1103/PhysRevC.94.024919} {\bibfield  {journal} {\bibinfo
  {journal} {Phys. Rev.}\ }\textbf {\bibinfo {volume} {C94}},\ \bibinfo {pages}
  {024919} (\bibinfo {year} {2016})}\BibitemShut {NoStop}%
\bibitem [{\citenamefont {Dusling}\ \emph
  {et~al.}(2017{\natexlab{a}})\citenamefont {Dusling}, \citenamefont {Mace},\
  and\ \citenamefont {Venugopalan}}]{Dusling:2017aot}%
  \BibitemOpen
  \bibfield  {author} {\bibinfo {author} {\bibfnamefont {K.}~\bibnamefont
  {Dusling}}, \bibinfo {author} {\bibfnamefont {M.}~\bibnamefont {Mace}}, \
  and\ \bibinfo {author} {\bibfnamefont {R.}~\bibnamefont {Venugopalan}},\
  }\href@noop {} {\  (\bibinfo {year} {2017}{\natexlab{a}})},\ \Eprint
  {http://arxiv.org/abs/1706.06260} {arXiv:1706.06260 [hep-ph]} \BibitemShut
  {NoStop}%
\bibitem [{\citenamefont {Dusling}\ \emph
  {et~al.}(2017{\natexlab{b}})\citenamefont {Dusling}, \citenamefont {Mace},\
  and\ \citenamefont {Venugopalan}}]{Dusling:2017dqg}%
  \BibitemOpen
  \bibfield  {author} {\bibinfo {author} {\bibfnamefont {K.}~\bibnamefont
  {Dusling}}, \bibinfo {author} {\bibfnamefont {M.}~\bibnamefont {Mace}}, \
  and\ \bibinfo {author} {\bibfnamefont {R.}~\bibnamefont {Venugopalan}},\
  }\href@noop {} {\  (\bibinfo {year} {2017}{\natexlab{b}})},\ \Eprint
  {http://arxiv.org/abs/1705.00745} {arXiv:1705.00745 [hep-ph]} \BibitemShut
  {NoStop}%
\bibitem [{\citenamefont {M{\"a}ntysaari}\ \emph {et~al.}(2017)\citenamefont
  {M{\"a}ntysaari}, \citenamefont {Schenke}, \citenamefont {Shen},\ and\
  \citenamefont {Tribedy}}]{Mantysaari:2017cni}%
  \BibitemOpen
  \bibfield  {author} {\bibinfo {author} {\bibfnamefont {H.}~\bibnamefont
  {M{\"a}ntysaari}}, \bibinfo {author} {\bibfnamefont {B.}~\bibnamefont
  {Schenke}}, \bibinfo {author} {\bibfnamefont {C.}~\bibnamefont {Shen}}, \
  and\ \bibinfo {author} {\bibfnamefont {P.}~\bibnamefont {Tribedy}},\
  }\href@noop {} {\  (\bibinfo {year} {2017})},\ \Eprint
  {http://arxiv.org/abs/1705.03177} {arXiv:1705.03177 [nucl-th]} \BibitemShut
  {NoStop}%
\bibitem [{\citenamefont {Ruiz~Arriola}\ and\ \citenamefont
  {Broniowski}(2017)}]{RuizArriola:2016ihz}%
  \BibitemOpen
  \bibfield  {author} {\bibinfo {author} {\bibfnamefont {E.}~\bibnamefont
  {Ruiz~Arriola}}\ and\ \bibinfo {author} {\bibfnamefont {W.}~\bibnamefont
  {Broniowski}},\ }\href {\doibase 10.1103/PhysRevD.95.074030} {\bibfield
  {journal} {\bibinfo  {journal} {Phys. Rev.}\ }\textbf {\bibinfo {volume}
  {D95}},\ \bibinfo {pages} {074030} (\bibinfo {year} {2017})}\BibitemShut
  {NoStop}%
\bibitem [{\citenamefont {Albacete}\ and\ \citenamefont
  {Soto-Ontoso}(2017)}]{Albacete:2016pmp}%
  \BibitemOpen
  \bibfield  {author} {\bibinfo {author} {\bibfnamefont {J.~L.}\ \bibnamefont
  {Albacete}}\ and\ \bibinfo {author} {\bibfnamefont {A.}~\bibnamefont
  {Soto-Ontoso}},\ }\href {\doibase 10.1016/j.physletb.2017.04.055} {\bibfield
  {journal} {\bibinfo  {journal} {Phys. Lett.}\ }\textbf {\bibinfo {volume}
  {B770}},\ \bibinfo {pages} {149} (\bibinfo {year} {2017})},\ \Eprint
  {http://arxiv.org/abs/1605.09176} {arXiv:1605.09176 [hep-ph]} \BibitemShut
  {NoStop}%
\bibitem [{\citenamefont {Glauber}(1959)}]{Glauber}%
  \BibitemOpen
  \bibfield  {author} {\bibinfo {author} {\bibfnamefont {R.~J.}\ \bibnamefont
  {Glauber}},\ }\href@noop {} {\emph {\bibinfo {title} {{High Energy Collision
  Theory}}}},\ edited by\ \bibinfo {editor} {\bibfnamefont {W.~E.~B.}\
  \bibnamefont {et~al}},\ Vol.~\bibinfo {volume} {1}\ (\bibinfo  {publisher}
  {Interscience, New York},\ \bibinfo {year} {1959})\BibitemShut {NoStop}%
\bibitem [{\citenamefont {Di~Giacomo}\ and\ \citenamefont
  {Panagopoulos}(1992)}]{DiGiacomo:1992hhp}%
  \BibitemOpen
  \bibfield  {author} {\bibinfo {author} {\bibfnamefont {A.}~\bibnamefont
  {Di~Giacomo}}\ and\ \bibinfo {author} {\bibfnamefont {H.}~\bibnamefont
  {Panagopoulos}},\ }\href {\doibase 10.1016/0370-2693(92)91311-V} {\bibfield
  {journal} {\bibinfo  {journal} {Phys. Lett.}\ }\textbf {\bibinfo {volume}
  {B285}},\ \bibinfo {pages} {133} (\bibinfo {year} {1992})}\BibitemShut
  {NoStop}%
\bibitem [{\citenamefont {Kovner}\ and\ \citenamefont
  {Wiedemann}(2002)}]{PhysRevD66034031}%
  \BibitemOpen
  \bibfield  {author} {\bibinfo {author} {\bibfnamefont {A.}~\bibnamefont
  {Kovner}}\ and\ \bibinfo {author} {\bibfnamefont {U.~A.}\ \bibnamefont
  {Wiedemann}},\ }\href {\doibase 10.1103/PhysRevD.66.034031} {\bibfield
  {journal} {\bibinfo  {journal} {Phys. Rev. D}\ }\textbf {\bibinfo {volume}
  {66}},\ \bibinfo {pages} {034031} (\bibinfo {year} {2002})}\BibitemShut
  {NoStop}%
\bibitem [{\citenamefont {Sch{\"a}fer}\ and\ \citenamefont
  {Shuryak}(1998)}]{Schafer:1996wv}%
  \BibitemOpen
  \bibfield  {author} {\bibinfo {author} {\bibfnamefont {T.}~\bibnamefont
  {Sch{\"a}fer}}\ and\ \bibinfo {author} {\bibfnamefont {E.~V.}\ \bibnamefont
  {Shuryak}},\ }\href {\doibase 10.1103/RevModPhys.70.323} {\bibfield
  {journal} {\bibinfo  {journal} {Rev. Mod. Phys.}\ }\textbf {\bibinfo {volume}
  {70}},\ \bibinfo {pages} {323} (\bibinfo {year} {1998})},\ \Eprint
  {http://arxiv.org/abs/hep-ph/9610451} {arXiv:hep-ph/9610451 [hep-ph]}
  \BibitemShut {NoStop}%
\bibitem [{\citenamefont {Broniowski}\ \emph {et~al.}(2009)\citenamefont
  {Broniowski}, \citenamefont {Rybczy{\'n}ski},\ and\ \citenamefont
  {Bo{\.z}ek}}]{Broniowski:2007nz}%
  \BibitemOpen
  \bibfield  {author} {\bibinfo {author} {\bibfnamefont {W.}~\bibnamefont
  {Broniowski}}, \bibinfo {author} {\bibfnamefont {M.}~\bibnamefont
  {Rybczy{\'n}ski}}, \ and\ \bibinfo {author} {\bibfnamefont {P.}~\bibnamefont
  {Bo{\.z}ek}},\ }\href {\doibase 10.1016/j.cpc.2008.07.016} {\bibfield
  {journal} {\bibinfo  {journal} {Comput. Phys. Commun.}\ }\textbf {\bibinfo
  {volume} {180}},\ \bibinfo {pages} {69} (\bibinfo {year} {2009})}\BibitemShut
  {NoStop}%
\bibitem [{\citenamefont {Loizides}\ \emph {et~al.}(2015)\citenamefont
  {Loizides}, \citenamefont {Nagle},\ and\ \citenamefont
  {Steinberg}}]{Loizides:2014vua}%
  \BibitemOpen
  \bibfield  {author} {\bibinfo {author} {\bibfnamefont {C.}~\bibnamefont
  {Loizides}}, \bibinfo {author} {\bibfnamefont {J.}~\bibnamefont {Nagle}}, \
  and\ \bibinfo {author} {\bibfnamefont {P.}~\bibnamefont {Steinberg}},\ }\href
  {\doibase 10.1016/j.softx.2015.05.001} {\bibfield  {journal} {\bibinfo
  {journal} {SoftwareX}\ }\textbf {\bibinfo {volume} {1-2}},\ \bibinfo {pages}
  {13} (\bibinfo {year} {2015})},\ \Eprint {http://arxiv.org/abs/1408.2549}
  {arXiv:1408.2549 [nucl-ex]} \BibitemShut {NoStop}%
\bibitem [{\citenamefont {Bernhard}\ \emph {et~al.}(2016)\citenamefont
  {Bernhard}, \citenamefont {Moreland}, \citenamefont {Bass}, \citenamefont
  {Liu},\ and\ \citenamefont {Heinz}}]{Bernhard:2016tnd}%
  \BibitemOpen
  \bibfield  {author} {\bibinfo {author} {\bibfnamefont {J.~E.}\ \bibnamefont
  {Bernhard}}, \bibinfo {author} {\bibfnamefont {J.~S.}\ \bibnamefont
  {Moreland}}, \bibinfo {author} {\bibfnamefont {S.~A.}\ \bibnamefont {Bass}},
  \bibinfo {author} {\bibfnamefont {J.}~\bibnamefont {Liu}}, \ and\ \bibinfo
  {author} {\bibfnamefont {U.}~\bibnamefont {Heinz}},\ }\href {\doibase
  10.1103/PhysRevC.94.024907} {\bibfield  {journal} {\bibinfo  {journal} {Phys.
  Rev.}\ }\textbf {\bibinfo {volume} {C94}},\ \bibinfo {pages} {024907}
  (\bibinfo {year} {2016})},\ \Eprint {http://arxiv.org/abs/1605.03954}
  {arXiv:1605.03954 [nucl-th]} \BibitemShut {NoStop}%
\bibitem [{\citenamefont {Bo{\.z}ek}\ \emph {et~al.}(2016)\citenamefont
  {Bo{\.z}ek}, \citenamefont {Broniowski},\ and\ \citenamefont
  {Rybczy{\'n}ski}}]{Bozek:2016kpf}%
  \BibitemOpen
  \bibfield  {author} {\bibinfo {author} {\bibfnamefont {P.}~\bibnamefont
  {Bo{\.z}ek}}, \bibinfo {author} {\bibfnamefont {W.}~\bibnamefont
  {Broniowski}}, \ and\ \bibinfo {author} {\bibfnamefont {M.}~\bibnamefont
  {Rybczy{\'n}ski}},\ }\href {\doibase 10.1103/PhysRevC.94.014902} {\bibfield
  {journal} {\bibinfo  {journal} {Phys. Rev.}\ }\textbf {\bibinfo {volume}
  {C94}},\ \bibinfo {pages} {014902} (\bibinfo {year} {2016})}\BibitemShut
  {NoStop}%
\bibitem [{\citenamefont {Loizides}(2016)}]{Loizides:2016djv}%
  \BibitemOpen
  \bibfield  {author} {\bibinfo {author} {\bibfnamefont {C.}~\bibnamefont
  {Loizides}},\ }\href {\doibase 10.1103/PhysRevC.94.024914} {\bibfield
  {journal} {\bibinfo  {journal} {Phys. Rev.}\ }\textbf {\bibinfo {volume}
  {C94}},\ \bibinfo {pages} {024914} (\bibinfo {year} {2016})},\ \Eprint
  {http://arxiv.org/abs/1603.07375} {arXiv:1603.07375 [nucl-ex]} \BibitemShut
  {NoStop}%
\bibitem [{\citenamefont {Mitchell}\ \emph {et~al.}(2016)\citenamefont
  {Mitchell}, \citenamefont {Perepelitsa}, \citenamefont {Tannenbaum},\ and\
  \citenamefont {Stankus}}]{Mitchell:2016jio}%
  \BibitemOpen
  \bibfield  {author} {\bibinfo {author} {\bibfnamefont {J.~T.}\ \bibnamefont
  {Mitchell}}, \bibinfo {author} {\bibfnamefont {D.~V.}\ \bibnamefont
  {Perepelitsa}}, \bibinfo {author} {\bibfnamefont {M.~J.}\ \bibnamefont
  {Tannenbaum}}, \ and\ \bibinfo {author} {\bibfnamefont {P.~W.}\ \bibnamefont
  {Stankus}},\ }\href {\doibase 10.1103/PhysRevC.93.054910} {\bibfield
  {journal} {\bibinfo  {journal} {Phys. Rev.}\ }\textbf {\bibinfo {volume}
  {C93}},\ \bibinfo {pages} {054910} (\bibinfo {year} {2016})},\ \Eprint
  {http://arxiv.org/abs/1603.08836} {arXiv:1603.08836 [nucl-ex]} \BibitemShut
  {NoStop}%
\bibitem [{\citenamefont {Denicol}\ \emph {et~al.}(2014)\citenamefont
  {Denicol}, \citenamefont {Gale}, \citenamefont {Jeon}, \citenamefont
  {Paquet},\ and\ \citenamefont {Schenke}}]{Denicol:2014ywa}%
  \BibitemOpen
  \bibfield  {author} {\bibinfo {author} {\bibfnamefont {G.~S.}\ \bibnamefont
  {Denicol}}, \bibinfo {author} {\bibfnamefont {C.}~\bibnamefont {Gale}},
  \bibinfo {author} {\bibfnamefont {S.}~\bibnamefont {Jeon}}, \bibinfo {author}
  {\bibfnamefont {J.~F.}\ \bibnamefont {Paquet}}, \ and\ \bibinfo {author}
  {\bibfnamefont {B.}~\bibnamefont {Schenke}},\ }\href@noop {} {\  (\bibinfo
  {year} {2014})},\ \Eprint {http://arxiv.org/abs/1406.7792} {arXiv:1406.7792
  [nucl-th]} \BibitemShut {NoStop}%
\bibitem [{\citenamefont {Alvioli}\ \emph {et~al.}(2009)\citenamefont
  {Alvioli}, \citenamefont {Drescher},\ and\ \citenamefont
  {Strikman}}]{Alvioli:2009ab}%
  \BibitemOpen
  \bibfield  {author} {\bibinfo {author} {\bibfnamefont {M.}~\bibnamefont
  {Alvioli}}, \bibinfo {author} {\bibfnamefont {H.~J.}\ \bibnamefont
  {Drescher}}, \ and\ \bibinfo {author} {\bibfnamefont {M.}~\bibnamefont
  {Strikman}},\ }\href {\doibase 10.1016/j.physletb.2009.08.067} {\bibfield
  {journal} {\bibinfo  {journal} {Phys. Lett.}\ }\textbf {\bibinfo {volume}
  {B680}},\ \bibinfo {pages} {225} (\bibinfo {year} {2009})}\BibitemShut
  {NoStop}%
\bibitem [{\citenamefont {Blaizot}\ \emph {et~al.}(2014)\citenamefont
  {Blaizot}, \citenamefont {Broniowski},\ and\ \citenamefont
  {Ollitrault}}]{Blaizot:2014wba}%
  \BibitemOpen
  \bibfield  {author} {\bibinfo {author} {\bibfnamefont {J.-P.}\ \bibnamefont
  {Blaizot}}, \bibinfo {author} {\bibfnamefont {W.}~\bibnamefont {Broniowski}},
  \ and\ \bibinfo {author} {\bibfnamefont {J.-Y.}\ \bibnamefont {Ollitrault}},\
  }\href {\doibase 10.1103/PhysRevC.90.034906} {\bibfield  {journal} {\bibinfo
  {journal} {Phys. Rev.}\ }\textbf {\bibinfo {volume} {C90}},\ \bibinfo {pages}
  {034906} (\bibinfo {year} {2014})}\BibitemShut {NoStop}%
\bibitem [{\citenamefont {Bierlich}\ \emph {et~al.}(2015)\citenamefont
  {Bierlich}, \citenamefont {Gustafson}, \citenamefont {L{\"o}nnblad},\ and\
  \citenamefont {Tarasov}}]{Bierlich:2014xba}%
  \BibitemOpen
  \bibfield  {author} {\bibinfo {author} {\bibfnamefont {C.}~\bibnamefont
  {Bierlich}}, \bibinfo {author} {\bibfnamefont {G.}~\bibnamefont {Gustafson}},
  \bibinfo {author} {\bibfnamefont {L.}~\bibnamefont {L{\"o}nnblad}}, \ and\
  \bibinfo {author} {\bibfnamefont {A.}~\bibnamefont {Tarasov}},\ }\href
  {\doibase 10.1007/JHEP03(2015)148} {\bibfield  {journal} {\bibinfo  {journal}
  {JHEP}\ }\textbf {\bibinfo {volume} {03}},\ \bibinfo {pages} {148} (\bibinfo
  {year} {2015})},\ \Eprint {http://arxiv.org/abs/1412.6259} {arXiv:1412.6259
  [hep-ph]} \BibitemShut {NoStop}%
\bibitem [{\citenamefont {Bialas}\ \emph {et~al.}(1977)\citenamefont {Bialas},
  \citenamefont {Czyz},\ and\ \citenamefont {Furmanski}}]{Bialas:1977en}%
  \BibitemOpen
  \bibfield  {author} {\bibinfo {author} {\bibfnamefont {A.}~\bibnamefont
  {Bialas}}, \bibinfo {author} {\bibfnamefont {W.}~\bibnamefont {Czyz}}, \ and\
  \bibinfo {author} {\bibfnamefont {W.}~\bibnamefont {Furmanski}},\ }\href@noop
  {} {\bibfield  {journal} {\bibinfo  {journal} {Acta Phys. Polon.}\ }\textbf
  {\bibinfo {volume} {B8}},\ \bibinfo {pages} {585} (\bibinfo {year}
  {1977})}\BibitemShut {NoStop}%
\bibitem [{\citenamefont {Bialas}\ \emph {et~al.}(1976)\citenamefont {Bialas},
  \citenamefont {Bleszynski},\ and\ \citenamefont {Czyz}}]{Bialas:1976ed}%
  \BibitemOpen
  \bibfield  {author} {\bibinfo {author} {\bibfnamefont {A.}~\bibnamefont
  {Bialas}}, \bibinfo {author} {\bibfnamefont {M.}~\bibnamefont {Bleszynski}},
  \ and\ \bibinfo {author} {\bibfnamefont {W.}~\bibnamefont {Czyz}},\ }\href
  {\doibase 10.1016/0550-3213(76)90329-1} {\bibfield  {journal} {\bibinfo
  {journal} {Nucl. Phys.}\ }\textbf {\bibinfo {volume} {B111}},\ \bibinfo
  {pages} {461} (\bibinfo {year} {1976})}\BibitemShut {NoStop}%
\bibitem [{\citenamefont {Cudell}\ \emph {et~al.}(2002)\citenamefont {Cudell},
  \citenamefont {Ezhela}, \citenamefont {Gauron}, \citenamefont {Kang},
  \citenamefont {Kuyanov}, \citenamefont {Lugovsky}, \citenamefont {Martynov},
  \citenamefont {Nicolescu}, \citenamefont {Razuvaev},\ and\ \citenamefont
  {Tkachenko}}]{Cudell:2002xe}%
  \BibitemOpen
  \bibfield  {author} {\bibinfo {author} {\bibfnamefont {J.~R.}\ \bibnamefont
  {Cudell}}, \bibinfo {author} {\bibfnamefont {V.~V.}\ \bibnamefont {Ezhela}},
  \bibinfo {author} {\bibfnamefont {P.}~\bibnamefont {Gauron}}, \bibinfo
  {author} {\bibfnamefont {K.}~\bibnamefont {Kang}}, \bibinfo {author}
  {\bibfnamefont {{\relax Yu}.~V.}\ \bibnamefont {Kuyanov}}, \bibinfo {author}
  {\bibfnamefont {S.~B.}\ \bibnamefont {Lugovsky}}, \bibinfo {author}
  {\bibfnamefont {E.}~\bibnamefont {Martynov}}, \bibinfo {author}
  {\bibfnamefont {B.}~\bibnamefont {Nicolescu}}, \bibinfo {author}
  {\bibfnamefont {E.~A.}\ \bibnamefont {Razuvaev}}, \ and\ \bibinfo {author}
  {\bibfnamefont {N.~P.}\ \bibnamefont {Tkachenko}} (\bibinfo {collaboration}
  {COMPETE}),\ }\href {\doibase 10.1103/PhysRevLett.89.201801} {\bibfield
  {journal} {\bibinfo  {journal} {Phys. Rev. Lett.}\ }\textbf {\bibinfo
  {volume} {89}},\ \bibinfo {pages} {201801} (\bibinfo {year}
  {2002})}\BibitemShut {NoStop}%
\bibitem [{\citenamefont {Aaboud}\ \emph {et~al.}(2016)\citenamefont {Aaboud}
  \emph {et~al.}}]{Aaboud:2016itf}%
  \BibitemOpen
  \bibfield  {author} {\bibinfo {author} {\bibfnamefont {M.}~\bibnamefont
  {Aaboud}} \emph {et~al.} (\bibinfo {collaboration} {ATLAS}),\ }\href
  {\doibase 10.1140/epjc/s10052-016-4335-y} {\bibfield  {journal} {\bibinfo
  {journal} {Eur. Phys. J.}\ }\textbf {\bibinfo {volume} {C76}},\ \bibinfo
  {pages} {502} (\bibinfo {year} {2016})}\BibitemShut {NoStop}%
\end{thebibliography}%
\bibliographystyle{apsrev4-1}
\end{document}